\documentclass[12pt]{iopart}

\usepackage{url}
\usepackage{graphicx}
\usepackage{amstext,amssymb}
\usepackage{xspace}
\usepackage{bbm}
\usepackage{multirow}
\usepackage{placeins}

\usepackage{algorithm}
\usepackage{algpseudocode}

\newcommand{\vectorspace}[2]{\ensuremath{\mathbb{#1}^{#2}}\xspace}
\newcommand{\point}[1]{\ensuremath{\vec{p_{#1}}}}
\newcommand{\set}[1]{\ensuremath{\mathbf{#1}}}

\newcommand{\mean}[1]{\ensuremath{\overline{#1}}}
\newcommand{\norm}[1]{\ensuremath{\left|\left|{#1}\right|\right|}}

\newcommand{\sublabel}[1]{\textbf{\footnotesize{(#1)}}}

\begin{document}

\title{Robust Structural Identification via Polyhedral Template Matching}
\author{Peter Mahler Larsen, S\o ren Schmidt, Jakob Schi\o tz}
\address{Department of Physics, Technical University of Denmark, 2800 Kgs. Lyngby, Denmark}
%\date{}

\begin{abstract}
Successful scientific applications of large-scale molecular dynamics often rely on automated methods for identifying the local crystalline structure of condensed phases.
Many existing methods for structural identification, such as Common Neighbour Analysis, rely on interatomic distances (or thresholds thereof) to classify atomic structure.  As a consequence they are sensitive to strain and thermal displacements, and preprocessing such as quenching or temporal averaging of the atomic positions is necessary to provide reliable identifications. We propose a new method, Polyhedral Template Matching (PTM), which classifies structures according to the topology of the local atomic environment, without any ambiguity in the classification, and with greater reliability than e.g. Common Neighbour Analysis in the presence of thermal fluctuations.  We demonstrate that the method can reliably be used to identify structures even in simulations near the melting point, and that it can identify the most common ordered alloy structures as well.  In addition, the method makes it easy to identify the local lattice orientation in polycrystalline samples, and to calculate the local strain tensor.  An implementation is made available under a Free and Open Source Software license.  
\end{abstract}
%\pacs{XXX}
\maketitle

\section{Introduction}

Often, the most challenging part of a Molecular Dynamics (MD) simulation is analysing the large amounts of data generated.  For simulations within condensed matter physics, algorithms for automatic determination of local structure are often necessary, for example to identify crystalline phases, defects in the crystal structure, or structural motifs in non-crystalline samples. One method is to use the energy of the individual atoms and identify defect atoms by a threshold value of the energy.  Another possibility is the centro-symmetry parameter, which is zero in many crystalline phases, but takes a non-zero value at defects such as dislocation cores~\cite{Kelchner:1998ev}.  Both methods are reasonably effective at identifying atoms near defects if the temperature is not too high, but do not provide additional information about the local structure.

A very popular analysis method is Common Neighbour Analysis (CNA)~\cite{Honeycutt:1987uo,Faken:1994kq}, which classifies bonds between atoms according the local bonding structure, and uses this to assign a local crystalline structure to the atoms.  The CNA has successfully been used to identify dislocations and grain boundaries in deforming polycrystalline metals~\cite{Schiotz:2003vu,Yamakov:2003hh,Brandl:2011bq}, local ordering in amorphous phases~\cite{Jonsson:1988ic,Bailey:2004ih}, and the competition between crystalline and icosahedral order in nanoparticles~\cite{Qi:2001jt,Wells:2015bb}.

The CNA method analyses the bonds between common neighbours of two bonded atoms.  It relies heavily on the concept of two atoms being ``bonded'' or ``neighbours'', and thus needs a strict definition of this; typically in the form of a cut-off distance defining if two atoms are neighbours.  Such a cut-off distance is, by necessity, somewhat arbitrary.  Furthermore, thermal vibrations, coexistence of various phases, and fluctuations in the local density may all influence the result and either introduce noise in the analysis, or make it impossible to find a cutoff parameter useful for the entire system.  For these reasons, Stukowski introduced the Adaptive Common Neighbour Analysis (ACNA)~\cite{Stukowski:2012ie}, where the cutoff distance is picked automatically, and individually, for each atom.  While this makes the method significantly more robust, the ACNA still suffers from noise introduced by thermal vibrations, causing bonds to be sporadically broken or formed.

Recently, Lazar \emph{et al.}\ have introduced a method classifying the local structure by the topology of the Voronoi cell surrounding the atom~\cite{Lazar:2015dva}.  While this makes the method more robust to thermal vibrations than the (A)CNA, the method has difficulty distinguishing between the face-centered cubic (fcc) and hexagonal closed packed (hcp) structures.%, and requires calculating the Voronoi cells of all atoms.

In this paper, we introduce the Polyhedral Template Matching (PTM), an approach which is similar in spirit to the method of Lazar \emph{et al.}.  The gist of the method is that the convex hull formed by the set of neighbouring atoms describes the local structure around an atom. The convex hull is represented as a planar graph, and this graph is then used to classify the structure.  As this method looks at a fixed number of neighbouring atoms around the atom being analysed, and as it does not employ a concept of bonds between these atoms, it is less sensitive to thermal fluctuations.  In addition, the method assigns a well-defined order to the neighbours of an atom, making it much easier to define a local orientation or a local elastic strain without referring to an initial unstrained structure.  An implementation of the method is available online~\cite{PTMgithubrepository}.

The structure of the paper is as follows:  in section~\ref{sec:template}, we review how structures can easily be evaluated against a template structure, given that mapping between the neighbouring atoms and the template atoms exists.  In section~\ref{sec:neighbour_ordering} we describe two methods for selecting the correct neighbouring atoms.  In section~\ref{sec:convexhulls} we show how the convex hull of the neighbouring atoms can be used to define a low number of candidate templates and mappings, and present the resulting algorithm for structure identification.  In section~\ref{sec:results}, we benchmark the algorithm.  In section~\ref{sec:alloys} we extend it to ordered alloys.  Finally, in sections~\ref{sec:orientations} and~\ref{sec:strain}, we illustrate how the method can provide information about local lattice orientation and local elastic strain, the former at no additional computational cost.

%%% Local Variables:
%%% mode: latex
%%% TeX-master: "article"
%%% End:

\section{Template Matching}
\label{sec:template}
In this section we will describe how template matching can be used to choose the best structural match, given that a point-to-point correspondence between simulated and template structures exists.  The process of finding the point-to-point correspondences is described in the next section.  A commonly used measure of similarity between two point sets is the Root-Mean-Square Deviation (RMSD).  Given two sets of points \set{v} and \set{w}, the RMSD is defined as:
\begin{equation}
\text{RMSD}(\set{v}, \set{w}) = \sqrt{\frac{1}{N} \sum\limits_{i=1}^{N}
\norm{
\vec{v_i} - \vec{w_i}
}_2^2}
\label{eq:rmsd_fixed}
\end{equation}

The superposition problem is that of finding a translation and a rotation of \set{w} and a scaling of \set{v} which minimizes the RMSD.  This is equivalent to Horn's scale-asymmetric formulation~\cite{Horn:1987hf}.  It can be shown that the optimal translation is equivalent to bringing the barycentre of each point set to the origin.  The optimal rotation and scaling are given by:
\begin{equation}
\text{RMSD}(\set{v}, \set{w}) = \min_{s, \set{Q}} \sqrt{\frac{1}{N} \sum\limits_{i=1}^{N}
\norm{
s[\vec{v_i} - \mean{\set{v}}]
-
\left(\set{Q} [\vec{w_i} - \mean{\set{w}}]^T\right)^T
}_2^2}
\label{eq:rmsd_superposition}
\end{equation}
where \set{Q} is a right-handed orthogonal matrix,
$\mean{\set{v}} = \frac{1}{N} \sum\limits_{i=1}^N \vec{v_i}$ and
$\mean{\set{w}} = \frac{1}{N} \sum\limits_{i=1}^N \vec{w_i}$ are the barycentres of \set{v} and \set{w} respectively, and $s$ is the optimal scaling of \set{v}.  Finding \set{Q} is a well studied problem with many different solution methods; Theobald~\cite{Theobald:2005cy} provides a good exposition of the problem.  Horn~\cite{Horn:1987hf} describes a solution for finding $s$ and shows that \set{Q} is independent of $s$.
We can make the RMSD scale invariant by scaling \set{w} such that the mean distance of each point from the origin (after translation) is 1:
\begin{equation}
\text{RMSD}(\set{v}, \set{w}) = \min_{s, \set{Q}} \sqrt{\frac{1}{N} \sum\limits_{i=1}^{N}
\norm{
s[\vec{v_i} - \mean{\set{v}}]
-
\frac{1}{l(\set{w})}\left(\set{Q} [\vec{w_i} - \mean{\set{w}}]^T\right)^T
}_2^2}
\label{eq:rmsd_scale_invariant}
\end{equation}
where:
\begin{equation}
l(\set{w}) = \frac{1}{N}\sum\limits_{i=1}^N ||\vec{w_i} - \mean{\set{w}}||
\label{eq:rmsd_scaling}
\end{equation}

\begin{table}[tb]
\centering
\begin{tabular}{|c|c|}
\hline
Structure & Neighbours Required\\
\hline
SC & 6\\
FCC & 12\\
HCP & 12\\
ICO & 12\\
BCC & 14\\
\hline
\end{tabular}
\caption{Number of neighbouring atoms required to identify structures.\label{table:neighbours}}
\end{table}
%
% \begin{table}[!h]
% \centering
% \begin{tabular}{|c|c|c|c|c|c|}
% \hline
% Structure & SC & FCC & HCP & ICO & BCC\\
% \hline
% Neighbours Required & 6 & 12 & 12 & 12 & 14\\
% \hline
% \end{tabular}
% \caption{Number of neighbouring atoms required to identify structures.\label{table:neighbours}}
% \end{table}
%

Suppose now that we want to determine the structure of a central atom and its neighbours.  We wish to determine if it has simple cubic (SC), face-centred cubic (FCC), hexagonal close packed (HCP), icosahedral (ICO) or body-centred cubic (BCC) structure.  For SC, FCC, HCP and ICO the positions of the first shell of neighbouring atoms are sufficient to identify the structure.  For BCC the first two shells are required.  Table~\ref{table:neighbours} shows the number of atoms required for each structure.  Correct identifying the shell to which a neighbouring atom belongs is nontrivial at high temperatures; this is discussed in section \ref{sec:neighbour_ordering}.
Given a set of reference templates corresponding to the atom positions of the aforementioned structures, the template which best matches an atom and its neighbours is the template which minimizes the RMSD after superposition.  In the above formulation, \set{v} contains the positions of the central atom and its neighbours and \set{w} contains those of the template.  A scale-invariant RMSD serves two purposes; it avoids preferential weighting of smaller templates and avoids the need for selecting bond lengths.

As we have demonstrated here, the task of structural identification would be simple if the optimal point-to-point correspondences were known.  Clearly, a brute force approach of testing all possible permutations of the neighbours is computationally infeasible.  Fast determination of point-to-point correspondences is the main contribution of this work; the algorithm for doing this is developed in section~\ref{sec:convexhulls}.

\section{Neighbouring Atom Shell Identification}
\label{sec:neighbour_ordering}
\begin{figure}[tp]
\centering
\begin{minipage}[t]{.45\textwidth}
  \centering
  \includegraphics[width=0.7\textwidth]{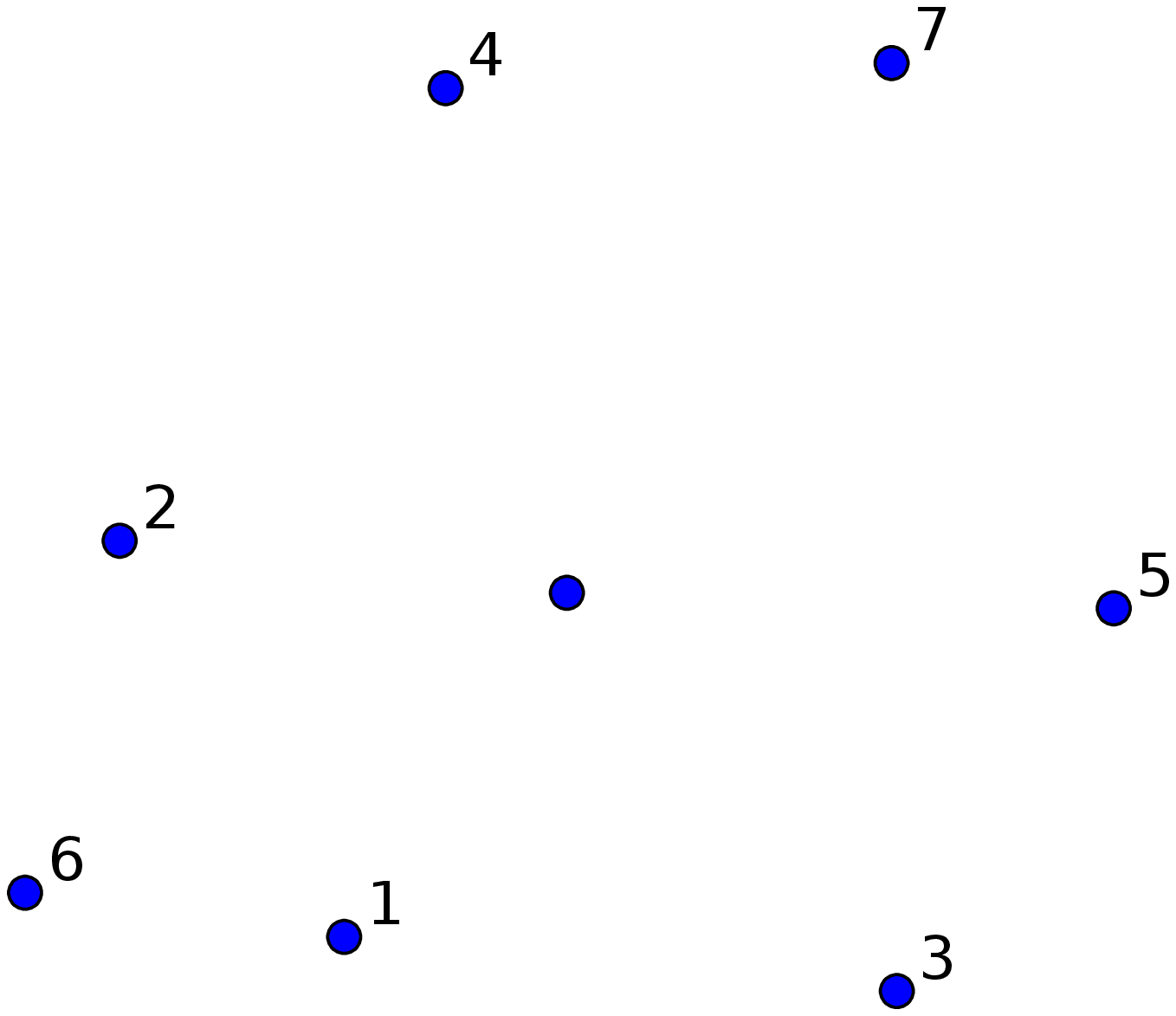}\\
  Euclidean ordering
\end{minipage}
\begin{minipage}[t]{.45\textwidth}
  \centering
  \includegraphics[width=0.7\textwidth]{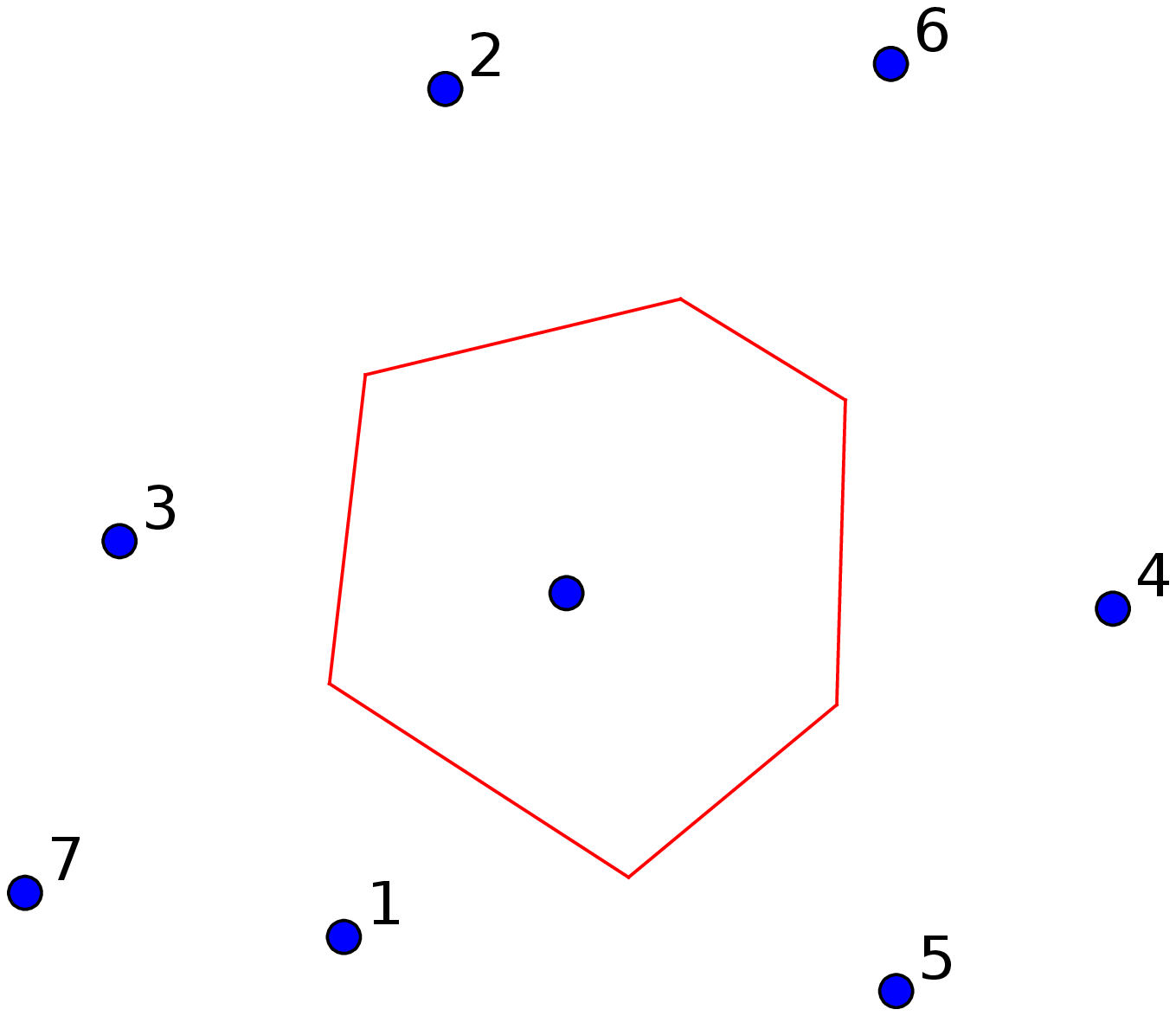}\\
  Topological ordering\\\phantom{ }
\end{minipage}
\caption{Neighbour atoms in a 2D hexagonal lattice ordered by Euclidean distance from the central atom (left) and by lexicographic ordering of the area of the polygonal interface between Voronoi cells and Euclidean distance (right).  A 2D example is used here for ease of illustration.  In \vectorspace{R}{2}, the `polygonal' interfaces between Voronoi cells are line segments, and the interfacial area of each is simply the length.}
\label{fig:neighbour_ordering}
\end{figure}

In order to use the template matching approach, we must correctly identify the nearest neighbours of each atom.  In a perfect lattice, all atoms in the same neighbour shell lie at the same distance from a central atom, by definition.  At low temperatures, thermal displacements are small enough such that the Euclidean distances from the central atom are sufficient to identify the shell numbers of neighbouring atoms.  At high temperatures, the thermal displacements are large enough that an atom in the second neighbour shell can be closer to the central atom than one in the first neighbour shell.  In this case, rather than using distance to order the neighbouring atoms, we can order them using the areas of the bounding polygons of the Voronoi cell of the central atom.  Given a discrete collection of points $\set{P} = \{\point{1}, \point{2}, \ldots, \point{N} \} \in \vectorspace{R}{3}$, the Voronoi cell of a point $p_i$ consists of all points in \vectorspace{R}{3} which are at least as close to $p_i$ as to any other $p_j$.  The boundary of a Voronoi cell can be defined by a set of polygons, each of which defines the interface to an adjacent Voronoi cell.  Since we also wish to order neighbouring atoms whose Voronoi cells are not adjacent to that of the central atom, we use a lexicographical ordering of the interfacial areas \emph{and} the distances:
\begin{equation}
\left( A_i, \frac{1}{d_i} \right) \geq \left( A_j, \frac{1}{d_j} \right) \hspace{10mm} \forall i < j
\label{eq:neighbour_ordering}
\end{equation}
where $A_i$ is the area of the polygonal interface between the Voronoi cells of the central atom and a neighbouring atom, and $d_i$ is the Euclidean distance between them.  In the case where the Voronoi cell of a neighbouring atom is not adjacent to that of the central atom, we assign an area of $A_i=0$.  We denote this ordering as the `topological' ordering, which is demonstrated for a 2D example in Figure \ref{fig:neighbour_ordering}.

%\FloatBarrier
\section{Convex Hulls}
\label{sec:convexhulls}
\begin{figure}[tp]
\centering
\begin{minipage}{.19\textwidth}
  \centering
  \includegraphics[width=1.0\textwidth]{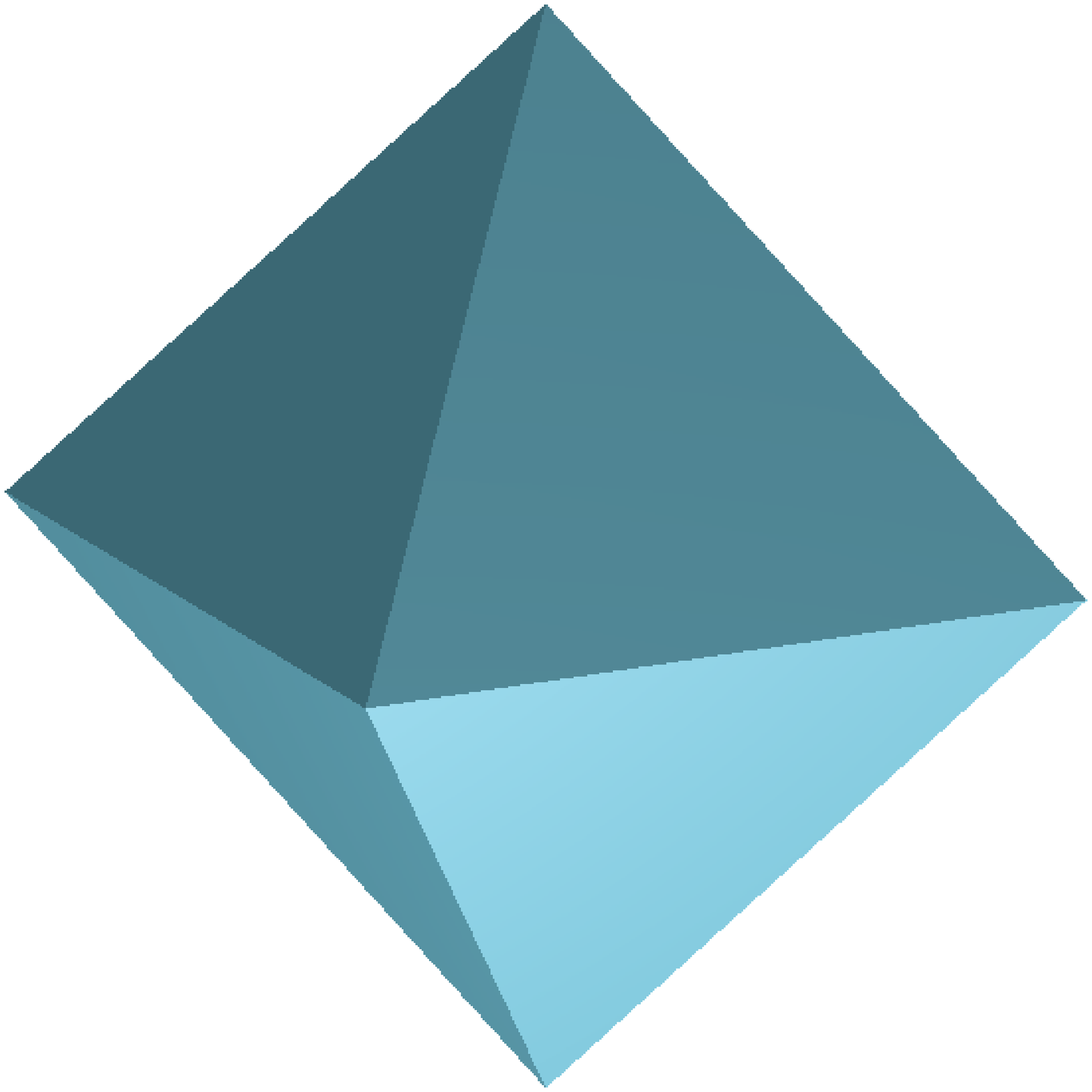}\\
  Simple Cubic (SC)
\end{minipage}
\begin{minipage}{.19\textwidth}
  \centering
  \includegraphics[width=1.0\textwidth]{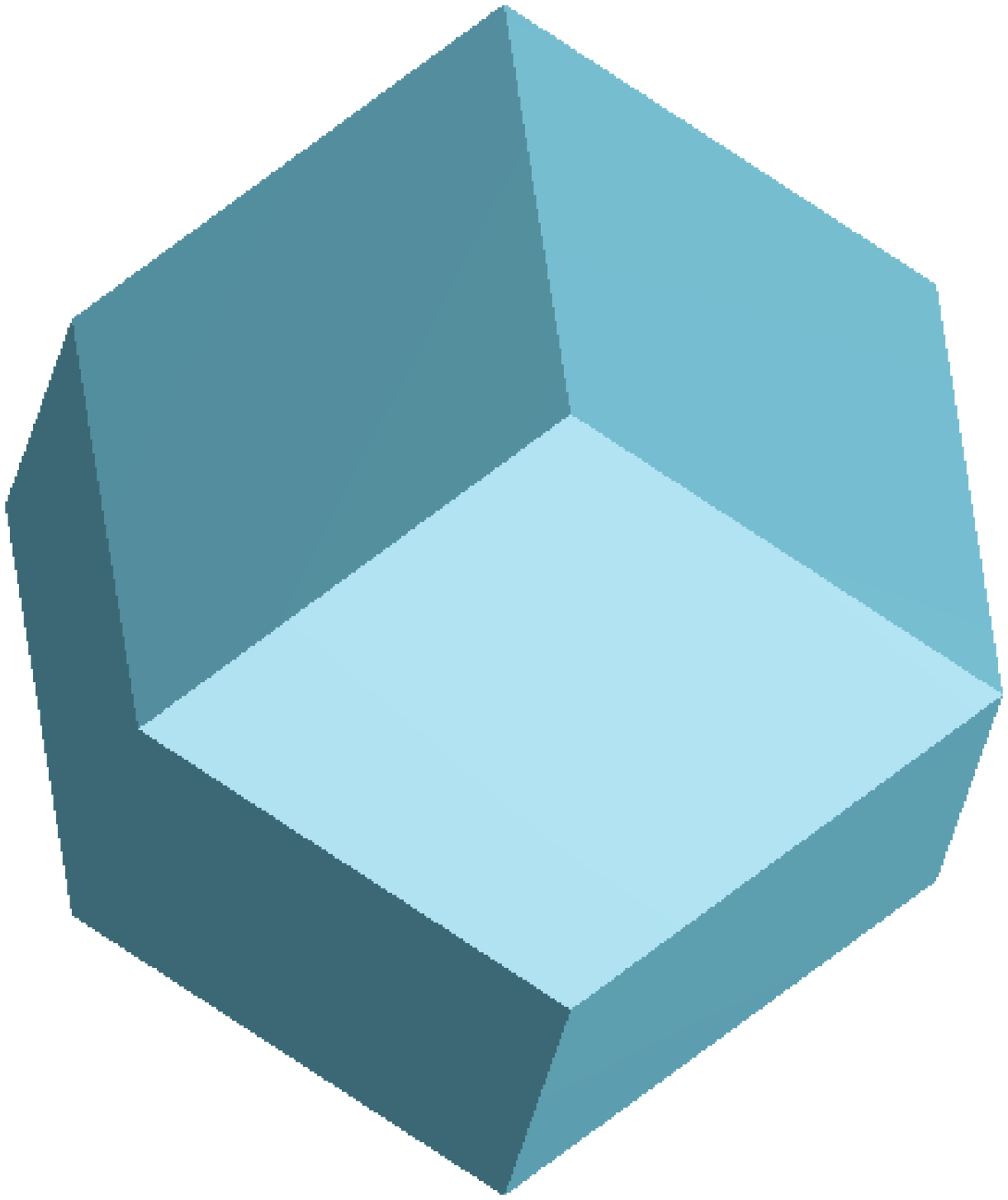}\\
  BCC\\\phantom{ }
\end{minipage}
\begin{minipage}{.19\textwidth}
  \centering
  \includegraphics[width=1.0\textwidth]{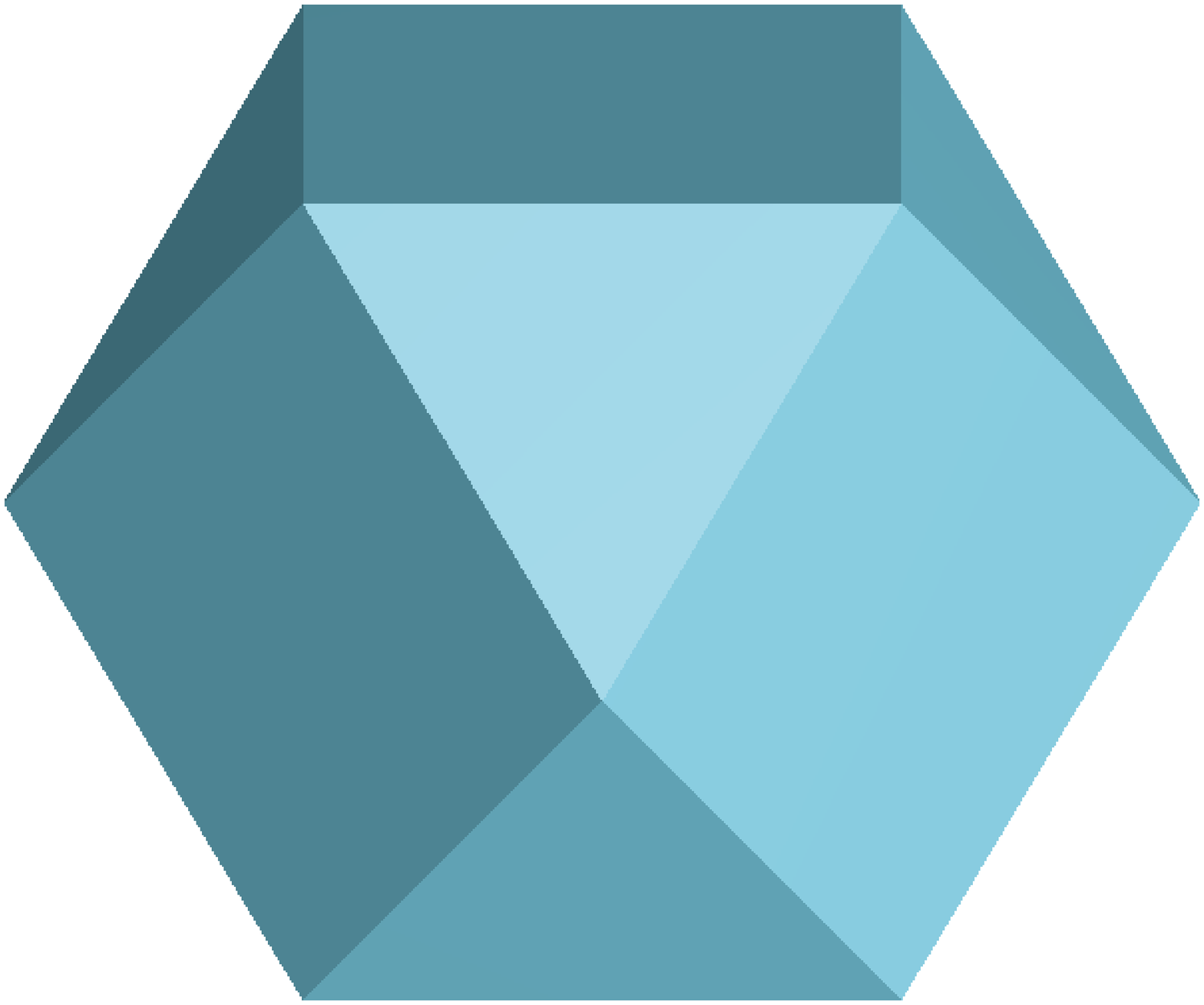}\\
  FCC\\\phantom{ }
\end{minipage}
\begin{minipage}{.19\textwidth}
  \centering
  \includegraphics[width=1.0\textwidth]{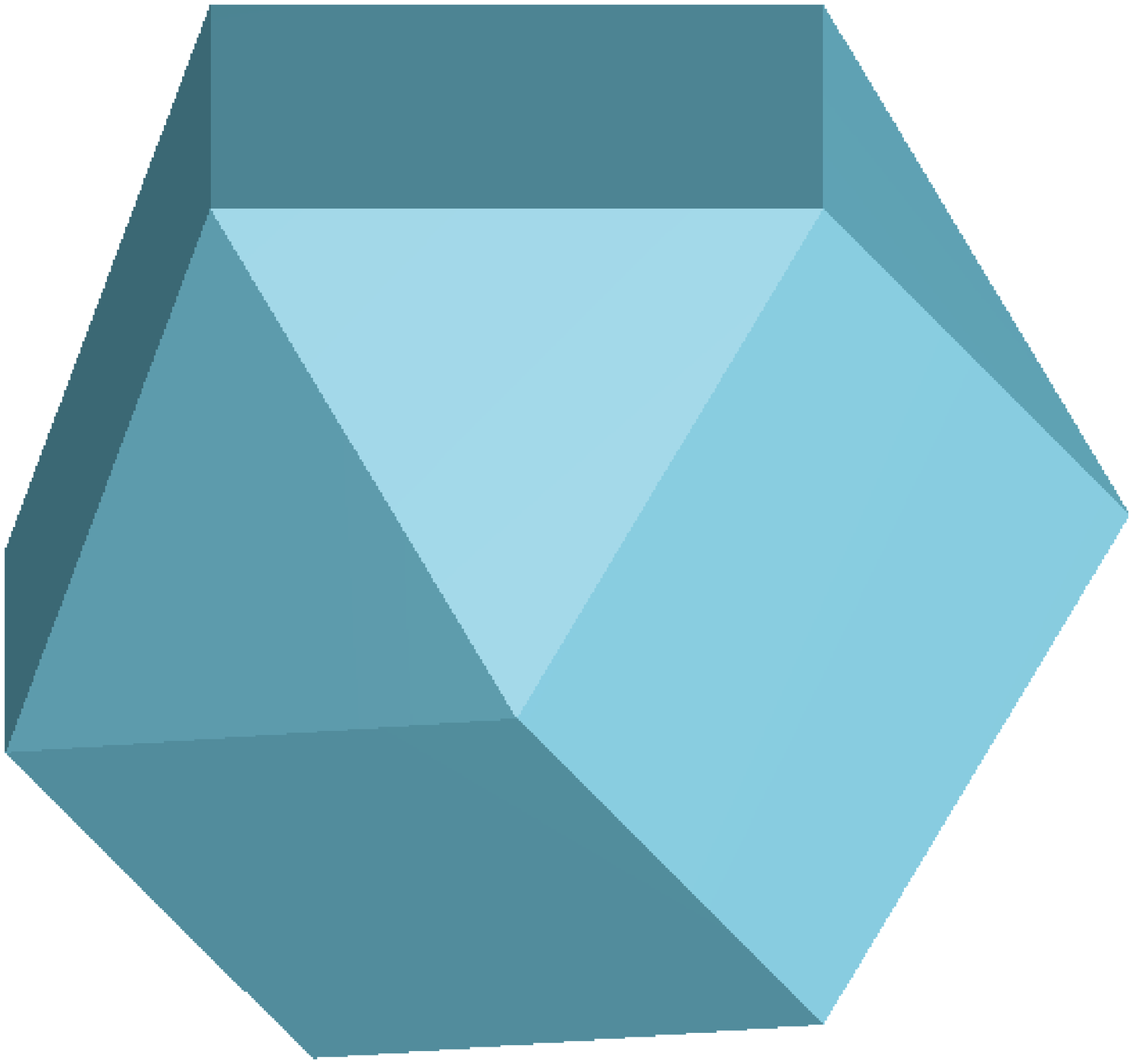}\\
  HCP\\\phantom{ }
\end{minipage}
\begin{minipage}{.19\textwidth}
  \centering
  \includegraphics[width=1.0\textwidth]{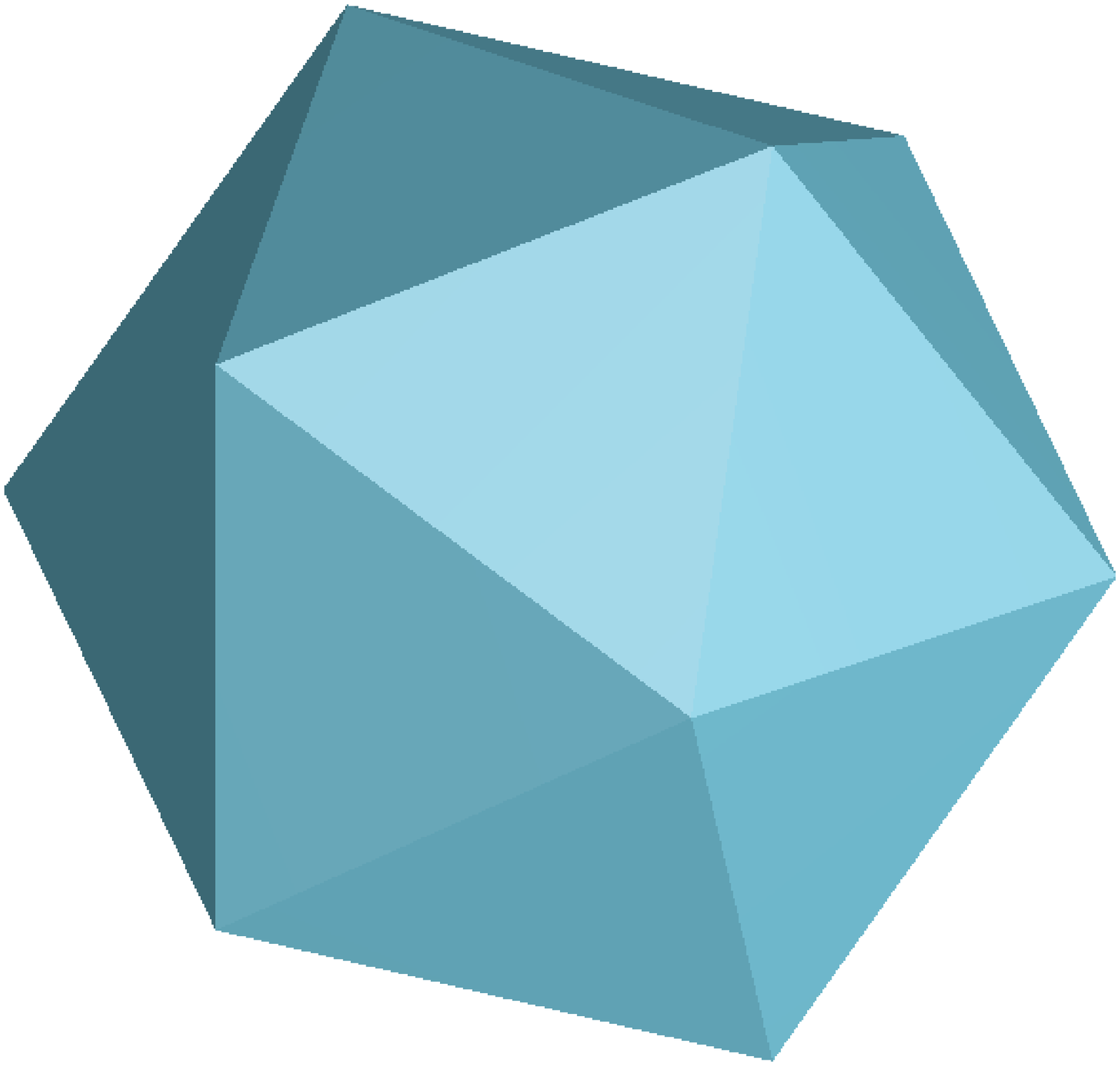}\\
  Icosahedral (ICO)
\end{minipage}
\caption{Convex hulls of the nearest neighbours of five different structures.  For the BCC structure, the first two shells of nearest neighbours are included.}
\label{fig:convex_hulls}
\end{figure}

We have described how template matching can be used to find the best structural match, given a known point-to-point correspondence.  In this section we will describe how convex hull graphs can be used find all possible point-to-point correspondences in an efficient manner.  A convex hull of a set of vertices is the smallest convex set which contains all the vertices.  In \vectorspace{R}{3} it can be described by a set of bounding planes.  Figure~\ref{fig:convex_hulls} shows the convex hulls of the five different structural types we wish to match.  The convex hull is the polyhedron formed by the nearest neighbours of a central atom.

\begin{figure}[htp]
\centering
\begin{minipage}{.19\textwidth}
  \centering
  \includegraphics[width=1.0\textwidth]{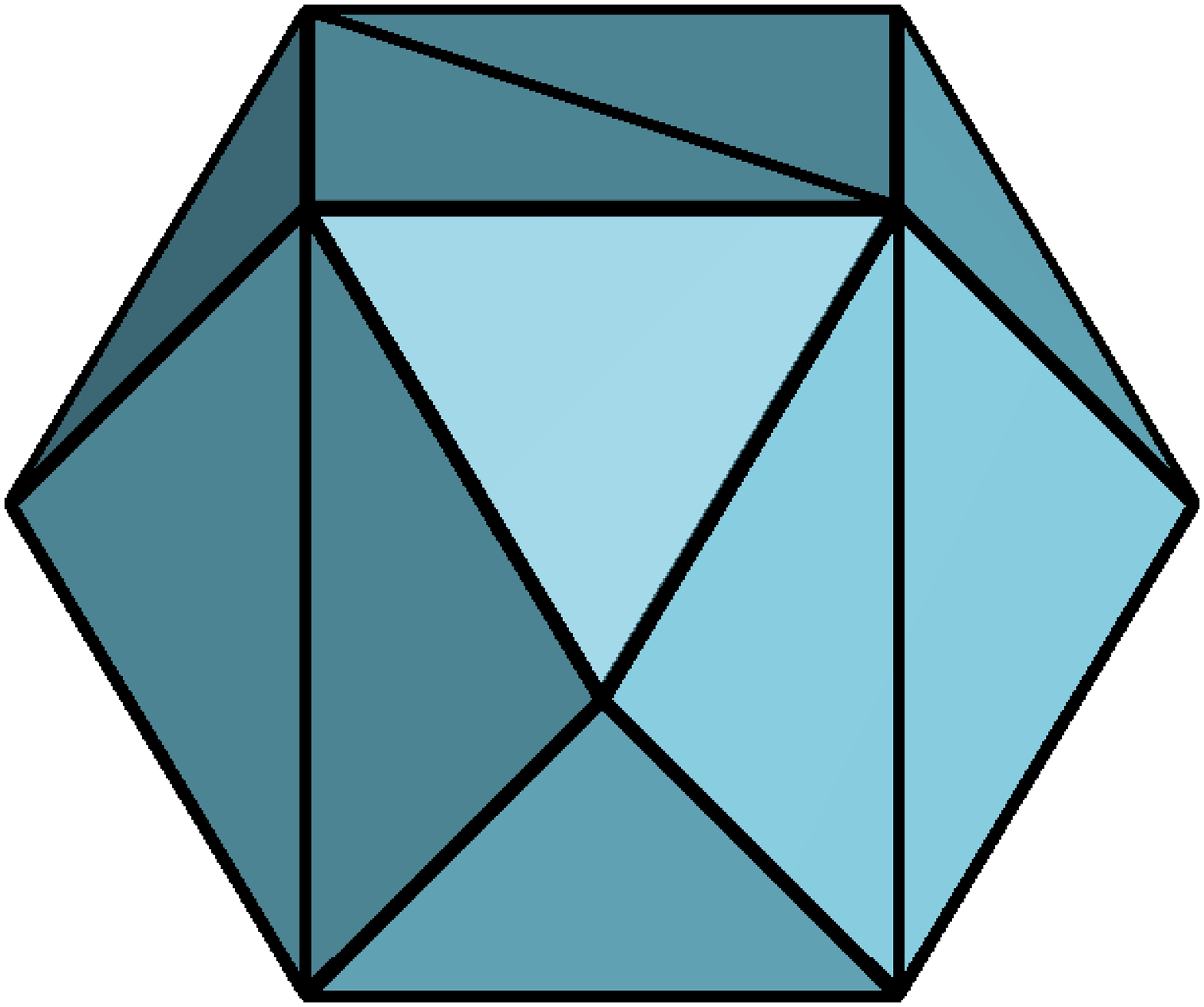}
\end{minipage}
\begin{minipage}{.19\textwidth}
  \centering
  \includegraphics[width=1.0\textwidth]{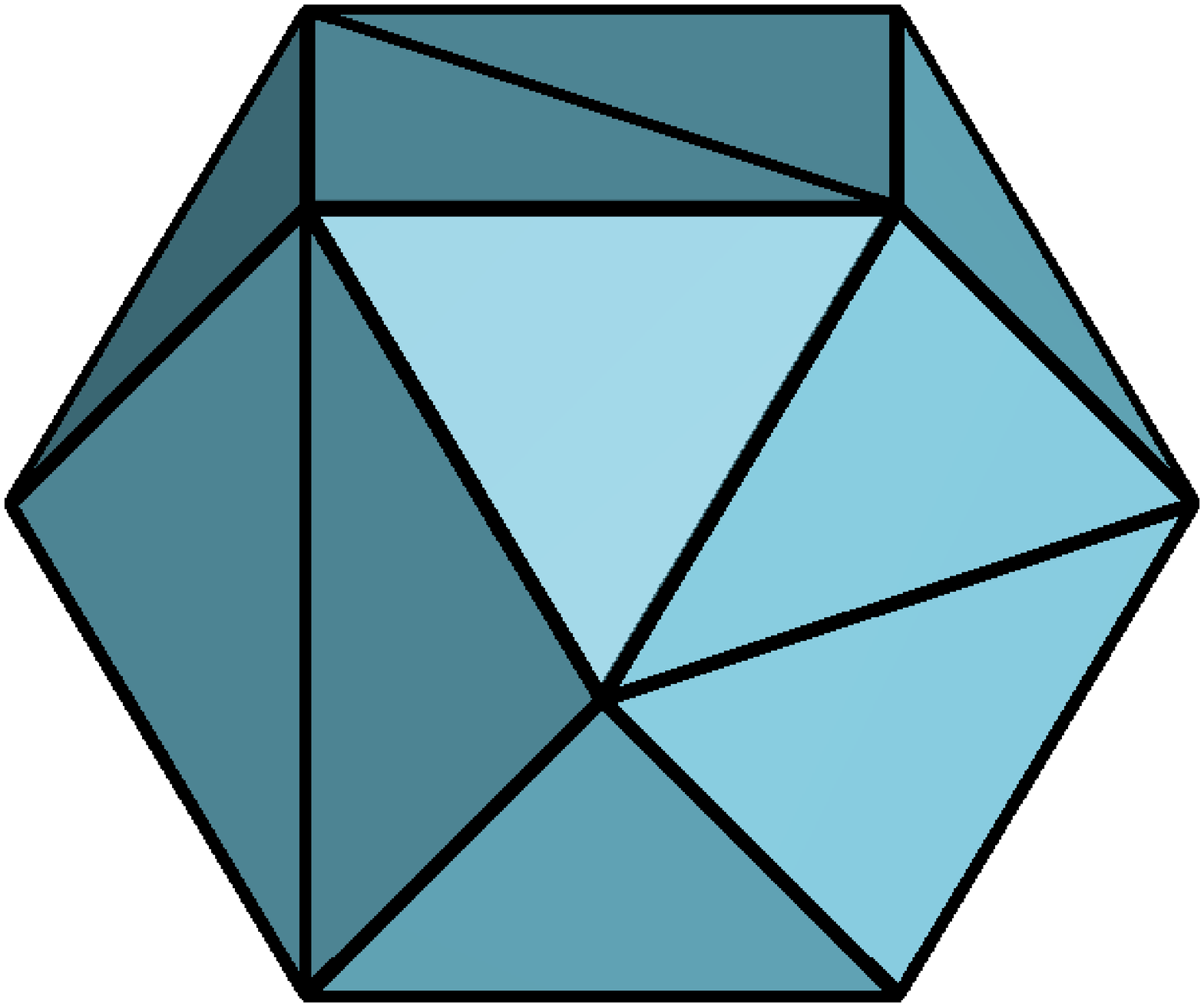}
\end{minipage}
\begin{minipage}{.19\textwidth}
  \centering
  \includegraphics[width=1.0\textwidth]{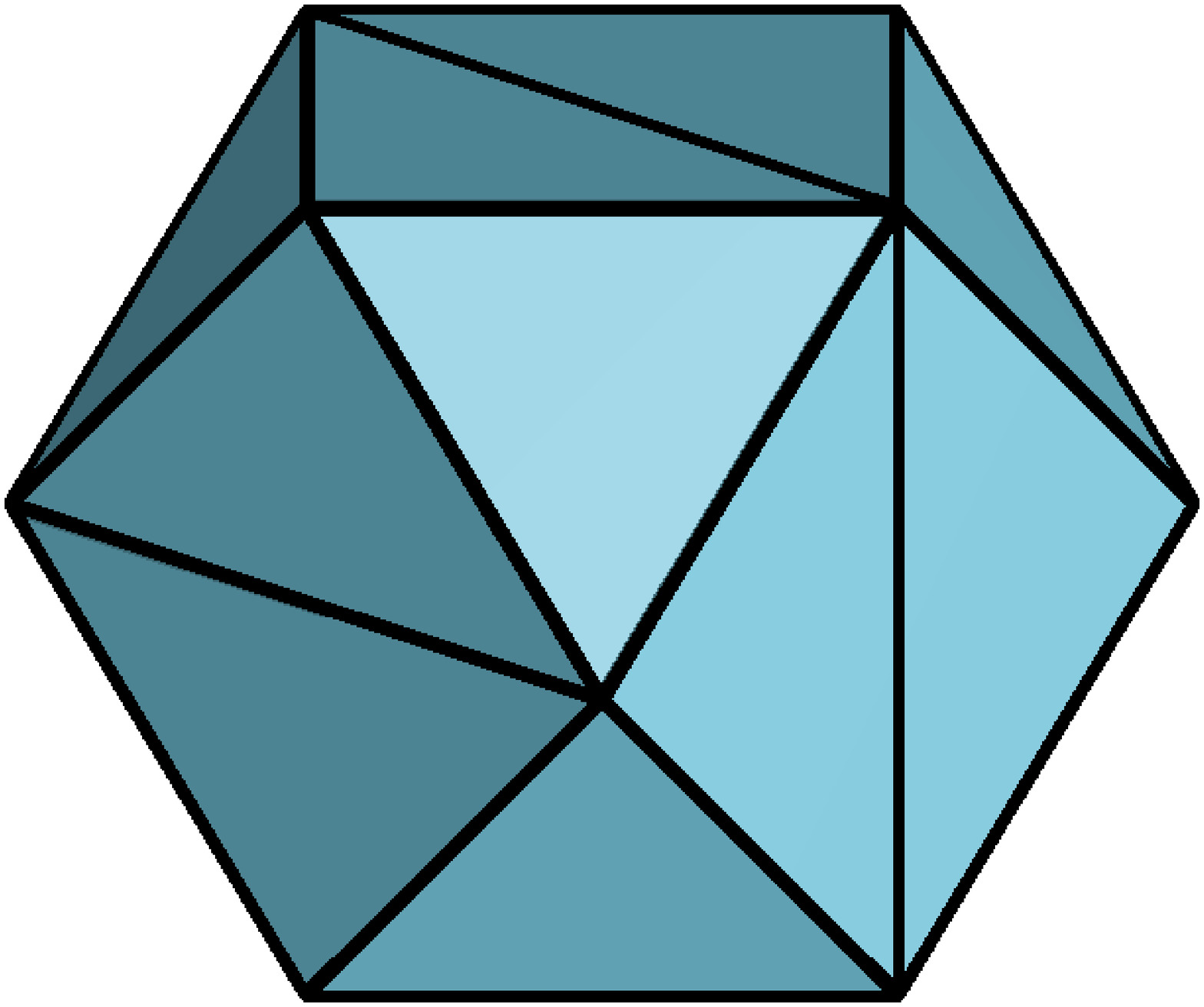}
\end{minipage}
\begin{minipage}{.19\textwidth}
  \centering
  \includegraphics[width=1.0\textwidth]{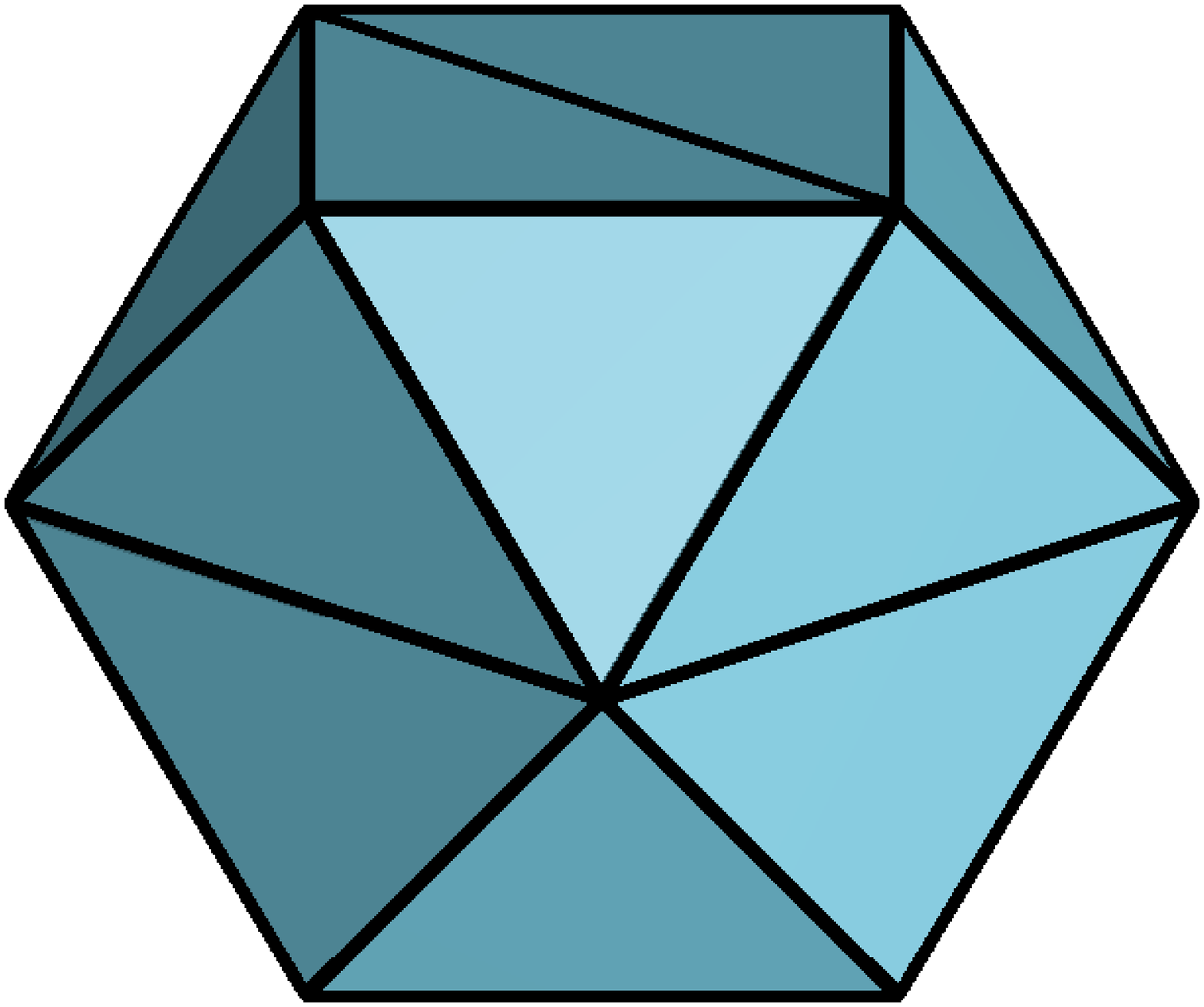}
\end{minipage}
\caption{Four different triangulations of the FCC convex hull. The convex hull contains six square facets, which gives a total of $2^6$ different triangulations.}
\label{fig:fcc_triangulations}
\end{figure}

Identification of the SC, FCC, HCP and ICO structures requires only the vertices of the first neighbour shell, which form a convex set.  Two neighbour shells are required for identification of the BCC structure but, remarkably, these vertices also form a convex set.  A stronger requirement of a convex hull is that it must consist only of simplicial facets, which in \vectorspace{R}{3} are triangles; this is a \emph{triangulation} of the surface of the convex hull.  We can furthermore require that the triangulations fulfil the empty-sphere condition, that is, no vertex is contained inside the circumcircle of another simplicial facet.  This is known as a \emph{Delaunay} triangulation.  Multiple Delaunay triangulations can exist for convex hulls with more than three coplanar points.
Let us first consider the SC and ICO cases.  The convex hulls of these structures consist only of triangular facets and the convex hull is therefore a unique triangulation.  The FCC and HCP convex hulls consist of both equilateral triangular facets and perfect square facets.  A perfect square has two equally valid triangulations which means the convex hull triangulation is not unique.  Furthermore, the small atomic displacements can change the triangulation of the square facets.  Figure~\ref{fig:fcc_triangulations} shows four different, equally valid, triangulations of the FCC convex hull.

The BCC convex hull consists only of rhombus facets.  Whilst an unperturbed rhombus facet has a unique Delaunay triangulation, small changes in vertex positions can change the triangulation.  We will consider both triangulations of each rhombus facet, since this will make our algorithm robust against relatively large atomic displacements.

\begin{figure}[htbp]
\centering
\begin{minipage}{.24\textwidth}
  \centering
  \includegraphics[width=0.9\textwidth]{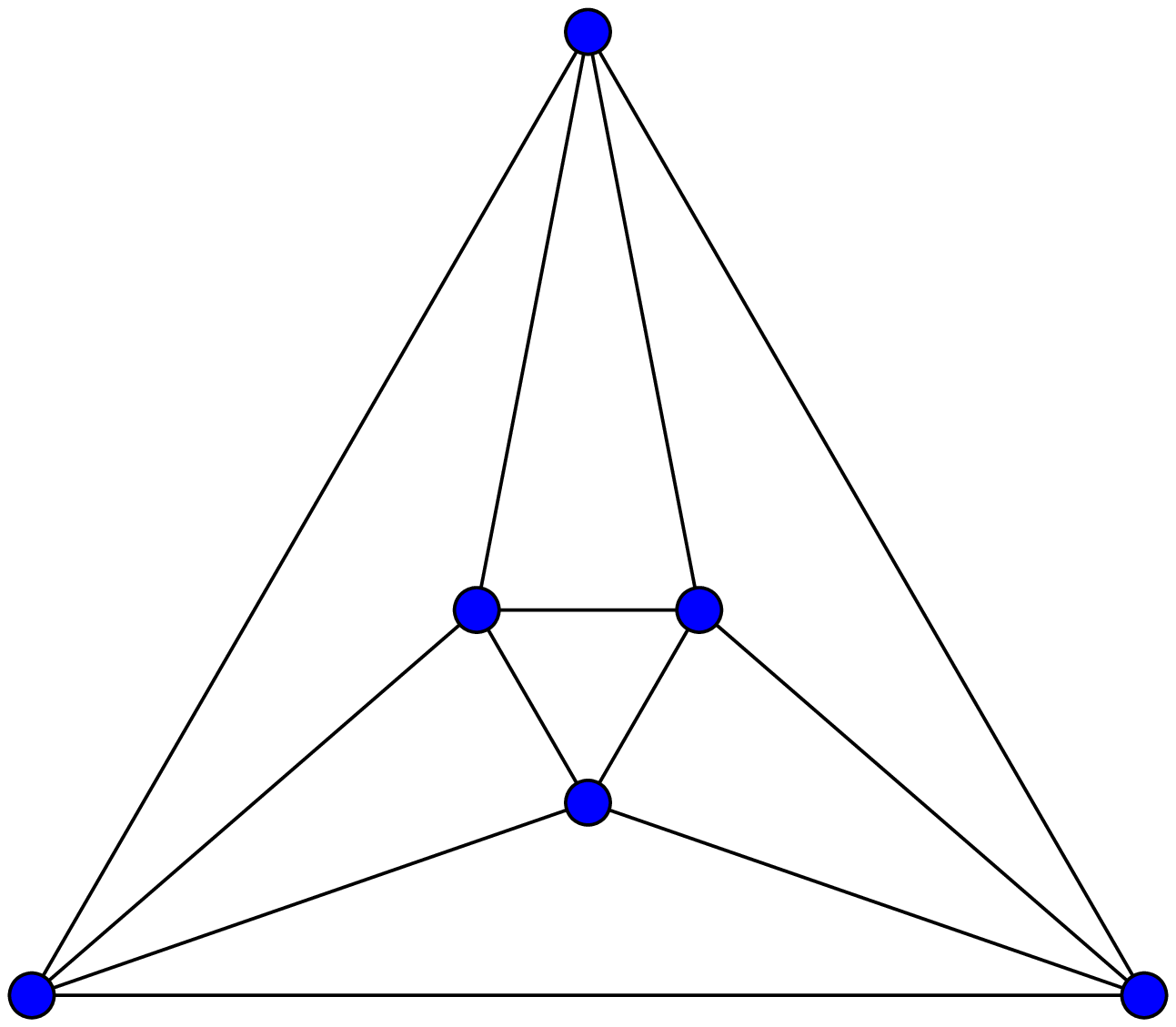}\\
  Simple Cubic
\end{minipage}
\begin{minipage}{.24\textwidth}
  \centering
  \includegraphics[width=0.9\textwidth]{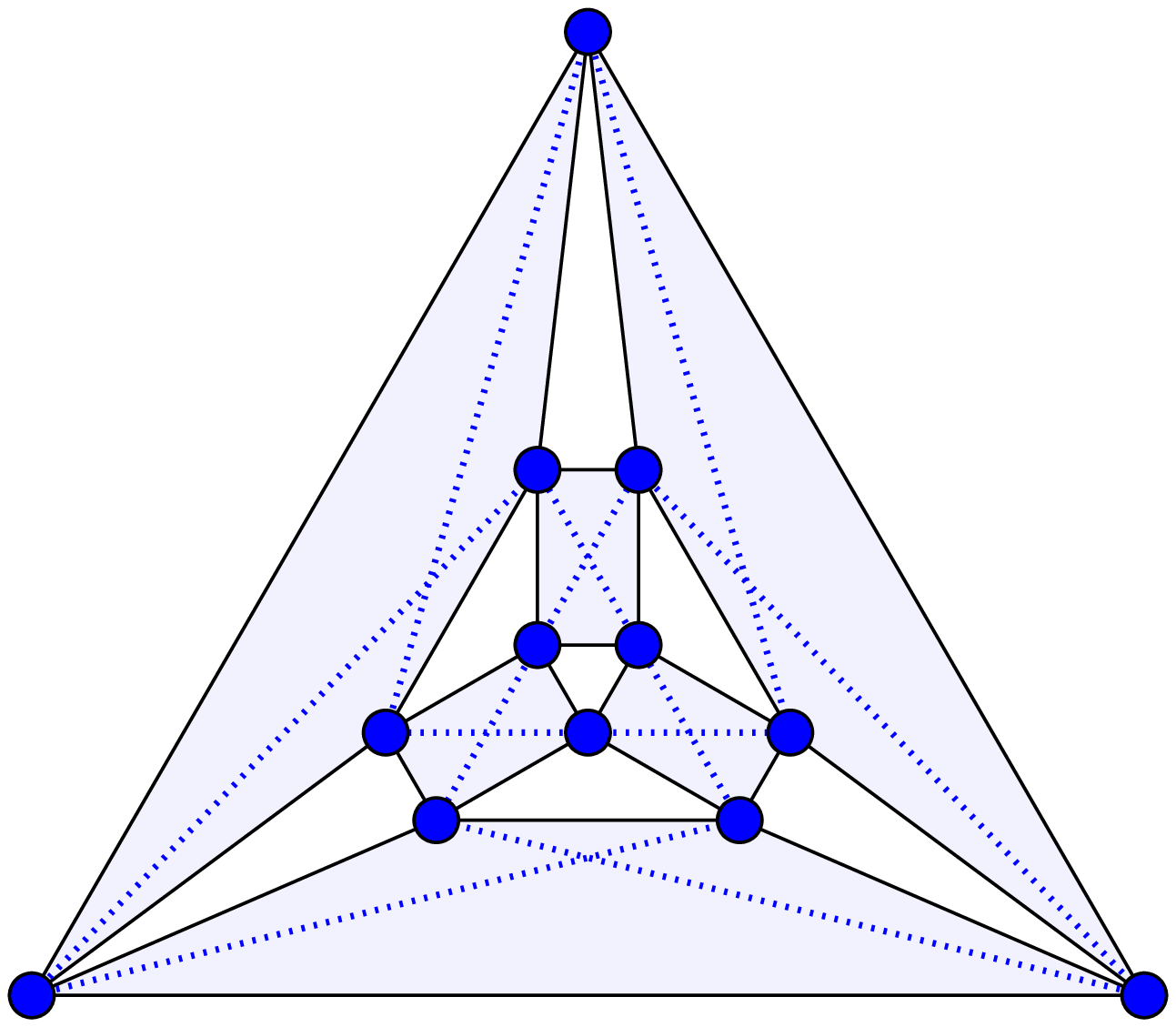}\\
  FCC
\end{minipage}
\begin{minipage}{.24\textwidth}
  \centering
  \includegraphics[width=0.9\textwidth]{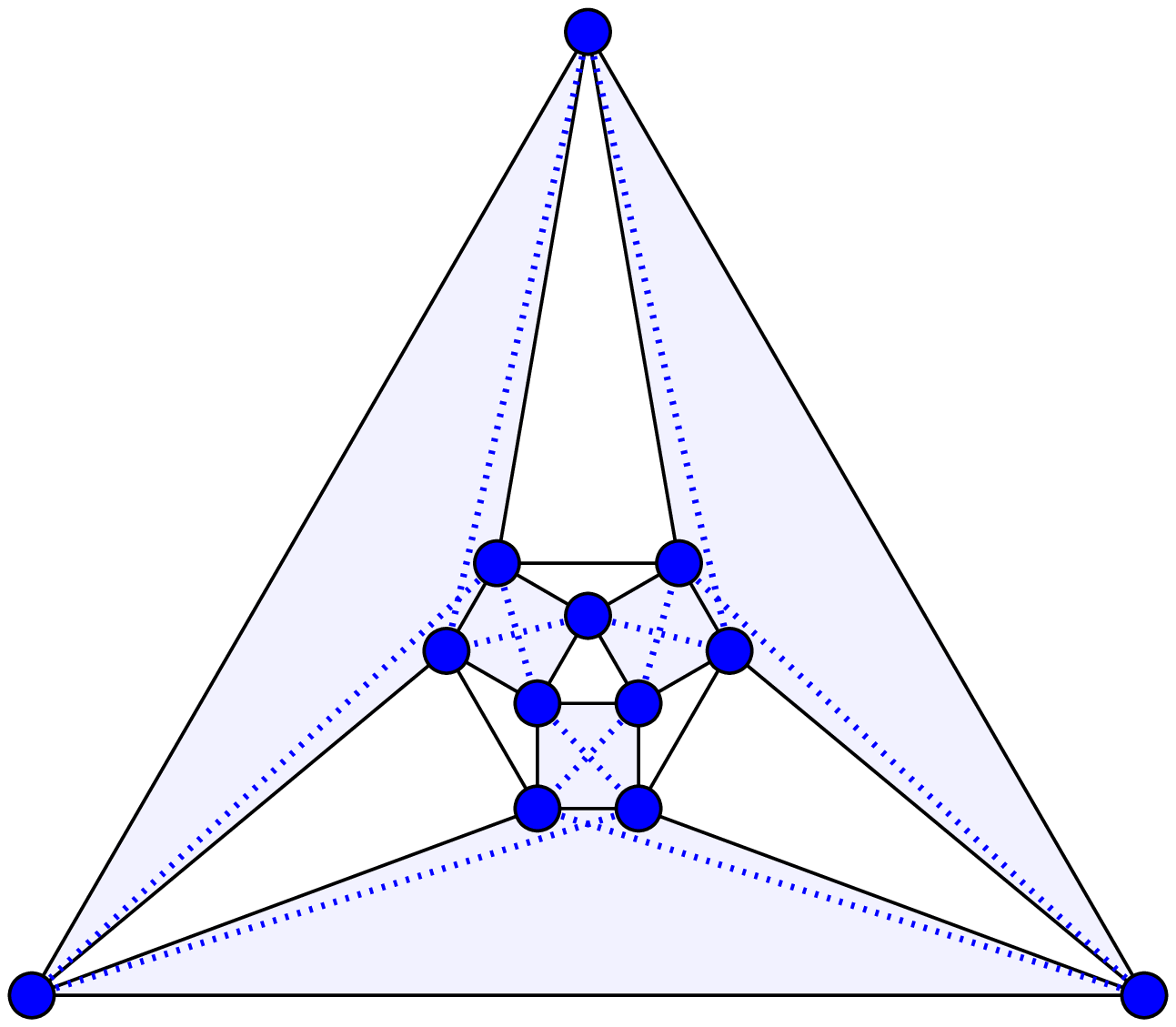}\\
  HCP
\end{minipage}
\begin{minipage}{.24\textwidth}
  \centering
  \includegraphics[width=0.9\textwidth]{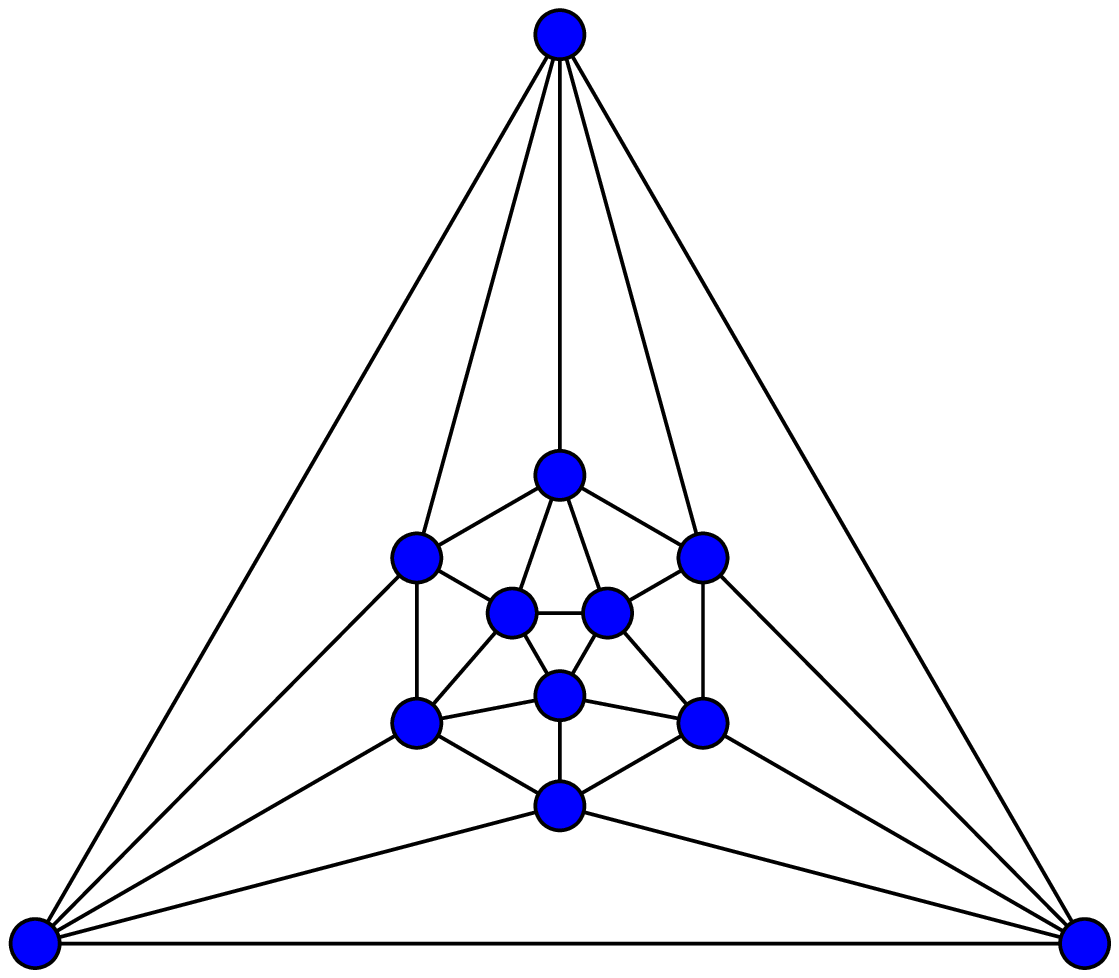}\\
  Icosahedral
\end{minipage}
\caption{Convex hulls in planar graph form.  The simple cubic and icosahedral graphs are unique, and are shown here as Tutte embeddings~\cite{Tutte:1963es}.  The FCC and HCP convex hulls do not have unique triangulations.  The regions of the planar graphs which correspond to the square facets have been shaded.  The dotted lines represent the two possible triangulations of each square facet. The BCC convex hull has no triangular facets, and cannot therefore be represented in the above manner.}
\label{fig:planar_graphs}
\end{figure}

We can represent the convex hull triangulations in graph form; vertices and edges in the convex hull triangulation correspond to vertices and edges in the graph.  By representing a convex hull as a graph, we move from a metric space to a purely topological space.  Steinitz's theorem~\cite{steinitz1916polyeder} states that the skeleton of a three-dimensional convex polyhedron is a tri-connected planar graph.  Conversely, any tri-connected planar graph has two embeddings in \vectorspace{R}{3}, i.e.\ it corresponds to two polyhedra that are mirror images of each other~\cite{Whitney:1933dp}.  Figure~\ref{fig:planar_graphs} shows the planar graph representations of the convex hulls shown in Figure~\ref{fig:convex_hulls}.  Since the FCC, HCP and BCC convex hull triangulations are not unique, neither are their planar graph representations.

We will use graph isomorphism to determine possible structure matches.  Two graphs $G = (V,E)$ and $H=(W,F)$ (where $V$ and $W$ are the graph vertices and $E$ and $F$ are the graph edges) are said to be \emph{isomorphic} if there exists a bijective function $f: V \rightarrow W$ such that for all $v, w \in V:\;
\{ v, w \} \in E \Leftrightarrow \{ f(v), f(w) \} \in F$.  An \emph{automorphism} is an isomorphism of $G$ to itself.  Many practical algorithms determine graph isomorphism by computing the \emph{canonical form} of each graph.  The canonical form of a graph $G$ is a uniquely defined automorphism of $G$ such that any two graphs are isomorphic if and only if their canonical forms are identical. Weinberg~\cite{Weinberg:1966jy} provides a simple method for defining the canonical forms and finding the automorphisms of tri-connected planar graphs (a more readable description is given by Kukluk et al.~\cite{Kukluk:2004bt}).  Weinberg's algorithm finds the canonical form by investigating both possible embeddings of a planar graph.  Since we are dealing with the graphs of polyhedra in \vectorspace{R}{3}, an embedding already exists.  We have therefore adapted the algorithm to investigate only a single embedding.  By doing so, we restrict graph isomorphism to \emph{orientation-preserving isomorphism}.  Orientation-preserving isomorphism extends the above definition of graph isomorphism (we use the notation of Brinkmann et al.~\cite{Brinkmann:2007fw}): two \emph{embedded} graphs are said to be orientation-preservingly isomorphic if there exists a bijective function $f_e: E \rightarrow F$ such that if $\{e_1, e_2,\ldots...,e_k\}$
is the set of edges incident with a vertex $v \in V$, in clockwise order, then $\{f_e(e_1), f_e(e_2), \ldots, f_e(e_k) \}$ is the set of edges incident with the vertex $f(v) \in W$, in clockwise order.  An \emph{orientation-preserving automorphism} is an orientation-preserving isomorphism of $G$ to itself.  Henceforth, when we refer to isomorphisms and automorphisms, the orientation preserving variants are implied.

Using the planar graphs we can rule out impossible structure matches: in order for a set of points to match a reference template, the planar graphs of their convex hulls must be isomorphic.  Unfortunately, the converse is not true; due to the multiple triangulations of the FCC and HCP convex hulls, the planar graphs are insufficient to identify structures uniquely.  There are triangulations of FCC and HCP convex hulls whose planar graphs are isomorphic, and triangulations of both which are isomorphic to the icosahedral planar graph.  As such, using planar graph isomorphism alone would sometimes lead to more than one matching structure.  Nevertheless, the planar graph representation gives us a set of point-to-point correspondences to investigate, one of which is the optimal correspondence.  We define the optimal correspondence as the one which minimizes the RMSD between a set of points and a reference template after superposition (as defined in Equation~(\ref{eq:rmsd_scale_invariant})).  The total number of possible correspondences to investigate for each reference template is the product of the number triangulations of the convex hull, the number of automorphisms of each planar graph, and the two embeddings of each planar graph in \vectorspace{R}{3}.  In practice, both the number of triangulations and automorphisms required is greatly reduced by the symmetries of each reference template (24-fold symmetry for simple cubic, BCC and FCC, 6-fold symmetry for HCP, 60-fold symmetry for icosahedral); the algorithm for generating the set of symmetrically inequivalent triangulations and automorphisms is described in~\ref{sec:app_unique}.

The structural identification process is described in Algorithm~\ref{alg:structure_proc}.  An outline of the algorithm is as follows: for each atom, we loop over the possible structures to identify. The convex hull formed by the neighbouring atoms is calculated (the number of neighbouring atoms depends on the candidate structure), and the canonical form of the corresponding graph is found.  We then loop over all possible triangulations of the reference structure (skipping triangulations that are symmetrically equivalent), and calculate the corresponding canonical form.  If the canonical form of the reference structure is identical to the canonical form of the actual complex hull, we have a \emph{possible} structural match.  The quality of that match is then tested.  This test is done by iterating over all symmetrically inequivalent automorphisms of the candidate graph, generating all possible structural templates which are then tested against the actual positions of the neighbouring atoms by optimizing the RMSD in Equation~(\ref{eq:rmsd_scale_invariant}).  The structure with the lowest RMSD is identified, and the algorithm returns the RMSD value, the corresponding orientation and the kind of structure identified.

%this is wrong, the rotation must be mapped into the fundamental zone.  Since both the triangulations of the reference structure and its automorphisms are reduced by the symmetry of the candidate structure, the corresponding rotation matrix lies within the fundamental zone of the symmetry group of the identified structure.
%
%
%\algdef{SE}[DOWHILE]{Do}{doWhile}{\algorithmicdo}[1]{\algorithmicwhile\ #1}%
\begin{algorithm}
\begin{algorithmic}
\Procedure{DetermineStructure}{$\set{A}$}
	\State \Comment{\set{A} is the set of positions of central atom and its nearest neighbours}
	\State \Comment{\set{A} is ordered using Euclidean or topological ordering (c.f. sec. \ref{sec:neighbour_ordering})}
	\State $r^* := \infty$ \Comment{RMSD of best match}
	%\State $\set{Q}^* := \{ 1, 0, 0, 0 \}$
	\State $\set{Q}^* := \set{1}$    \Comment{Unit rotation}
	%\State $\set{W}^* := \text{disordered}$
	\State $\set{S}^* := \text{disordered}$    \Comment{No structure identified (yet)}
%
	%\For {each reference template $\set{W}_i \in \{\set{W}_{\text{SC}}, %\set{W}_{\text{FCC}}, \set{W}_{\text{HCP}}, \set{W}_{\text{ICO}}, %\set{W}_{\text{BCC}} \}$}
    \For {$\set{S} \in \{\text{SC, FCC, HCP, ICO, BCC}\}$}
        \State $\set{W} := \text{ReferenceTemplate}(\set{S})$
		\State $\set{U} = \{ \vec{a_j} \in \set{A} \;\;|\;\; j \leq |\set{W}| \}$ \Comment{Select innermost atoms}

		\State $\set{C} := \text{Conv}(\set{U})$	\Comment{Calculate convex hull of \set{U}}
		\If {$\vec{a}_1 \notin \set{C}$}	\Comment{Convex hull must not contain central atom}
			\State $\set{G} := \text{CanonicalForm(Graph}(\set{C}))$
			\For {each triangulation $\set{T}_{i}$ of $\text{Conv}(\set{W})$}
				\State $\set{G}_\text{ref} := \text{CanonicalForm(Graph}(\set{T}_{i}))$
				\If {$\set{G} = \set{G}_{\text{ref}}$} \Comment{Test graph isomorphism}
					\For {each automorphism $\set{A}_j$ of $\set{G}$}
						\State $\set{U}^\prime := \set{U} \circ \set{A}_j$ \Comment{Permute by automorphism}
						\State $\{r, \set{Q}\} := \text{RMSD}(\set{U}^\prime, \set{W})$ \Comment{Optimal superposition (c.f. Eq.~\ref{eq:rmsd_scale_invariant})}
						\If {$r < r^*$}
							\State $r^* := r$
							\State $\set{Q}^*:= \set{Q}$
							\State $\set{S}^* := \set{S}$
						\EndIf
					\EndFor
				\EndIf
			\EndFor
		\EndIf
	\EndFor
	\State \Return $\{r^*, \set{Q}^*, \set{S}^* \}$
\EndProcedure
\end{algorithmic}
\caption{Pseudocode for determining local structure around a single atom.  As argument, the procedure takes the positions of the atom and its neighbours, sorted in distance from the central atom.  The algorithm returns the RMSD of the match found (or infinity if no match), the rotation matrix corresponding to the match, and the structure.}
\label{alg:structure_proc}
\end{algorithm}
%
%
%\FloatBarrier
\section{Results}
\label{sec:results}
\subsection{Copper Precipitate Benchmark}
\label{sec:results_benchmark}
To measure the capabilities of the method, we will first make a comparison with two existing structural identification methods, Adaptive Common Neighbour Analysis (ACNA) and Neighbour Distance Analysis (NDA), both due to Stukowski~\cite{Stukowski:2012ie}.  Stukowski uses a Cu-rich 9R precipitate in a BCC-Fe system with 88737 atoms in total as a benchmark to compare a number of structural classification methods.  The benchmarked system, which is publicly available~\cite{stukowksi2012benchmark}, is a case where CNA~\cite{Honeycutt:1987uo}, Bond-Angle Analysis~\cite{Ackland:2006kg} and Voronoi-analysis perform badly, and ACNA and NDA perform very well.

\begin{figure}[htb]
\centering
\includegraphics[width=0.9\textwidth]{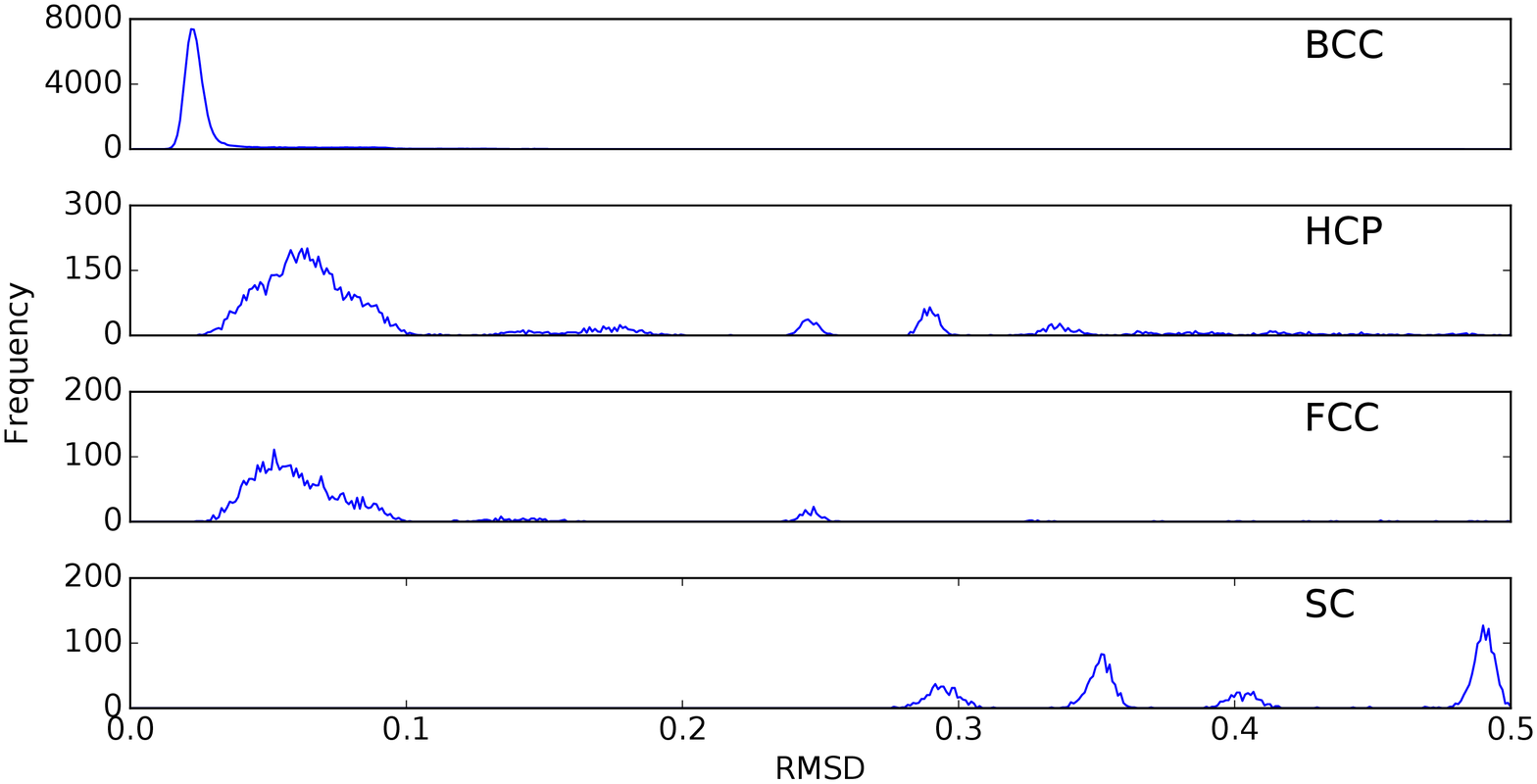}
\caption{RMSD histograms of BCC, HCP, FCC and SC structures in a Cu-rich 9R precipitate in BCC-Fe system.  The BCC atoms have very low RMSD values, which means the local structure is highly ordered.  The HCP and FCC structures have both slightly higher RMSD values which are nonetheless crystalline, and very high RMSD spurious identifications.  The SC identifications are exclusively spurious and are in fact highly disordered structures.}
\label{fig:hist_stukowksi_rmsd}
\end{figure}
We have applied our PTM algorithm to the same benchmark system.  Figure~\ref{fig:hist_stukowksi_rmsd} shows histograms of the RMSD values of the structures identified as BCC, HCP, FCC and SC.  We see that a large number of atoms are classified as BCC, FCC and HCP with relatively low RMSD, these are the atoms belonging to locally crystalline areas.  In addition, a much lower number of atoms are identified as HCP, FCC or SC but with a much higher RMSD.  These are atoms in locally disordered structures, but where the local environment provides a poor match to one of the structural templates.  One can choose to eliminate these spurious matches by introducing a cut-off, $\text{RMSD}_{\text{max}}$, for example based on a histogram such as Figure~\ref{fig:hist_stukowksi_rmsd}.

Table~\ref{table:stukowski_comparison} compares the performance of NDA and ACNA with PTM, both with no cut-off ($\text{RMSD}_{\text{max}} = \infty$) and with a sensible cut-off ($\text{RMSD}_{\text{max}} = 0.12$).  PTM is capable of indexing even highly distorted structures, which is very useful at high temperatures, but a good cut-off is required to avoid spurious classifications, such as the SC and ICO classifications shown in Table~\ref{table:stukowski_comparison}.   Other than plotting a RMSD-histogram, there is no simple method for selecting $\text{RMSD}_{\text{max}}$, but the cut-off can fortunately be chosen \emph{after} the analysis itself has been performed.  The `correct' value is dependent on the system being studied, as we will show in section~\ref{sec:results_cu3pt}.  For this reason, future publications which use PTM should report the value of $\text{RMSD}_{\text{max}}$ used to ensure reproducibility.

Note that the system here has been quenched, and as a consequence the local structures exhibit little distortion.  In this case PTM does not provide any real improvements in structural identification compared to ACNA and NDA.  The significant benefits of PTM are to be found in high-temperature systems.
\begin{table}[tb]
\centering
\begin{tabular}{|l|c|c|c|c|c|c|c|}
\hline
Analysis Method & Time (s)	& Disordered	& SC	& FCC	& HCP	& ICO	& BCC\\
\hline
$\text{ACNA}^\dagger$			& 0.66		& 12789			& 0		& 3138	& 7108	& 0		& 65702\\
$\text{NDA}^\dagger$				& 5.86		& 13260			& 0		& 3453	& 7825	& 0		& 64199\\
PTM $\text{RMSD}_{\text{max}}=0.12$	& 0.82	& 12900			& 0		& 3159	& 6971	& 0		& 65707\\
PTM $\text{RMSD}_{\text{max}}=\infty$	& 0.82	& 7115			& 2789	& 3420	& 8970	& 8		& 66435\\
\hline
\end{tabular}
\caption{Comparison of of PTM with two other structural analysis methods: Adaptive Common Neighbour Analysis (ACNA) and Neighbour Distance Analysis (NDA).  PTM gives similar results to both methods, though with a slightly longer running time than ACNA.  The neighbours are ordered by Euclidean distance rather than topologically since the system has been quenched.  A good $\text{RMSD}_{\text{max}}$ cut-off is required to avoid false positives. ${}^\dagger$Both the benchmark data and the implementations of ACNA and NDA are available online~\cite{stukowksi2012benchmark}.  Running times are for analysis only and do not include neighbour-list generation.  They were measured on a 2014 MacBook Pro with an Intel Core i7-4770HQ 2.20GHz CPU and 16GB RAM.
\label{table:stukowski_comparison}}
\end{table}

\subsection{Copper-Platinum Alloy System at High Temperature}
\label{sec:results_cu3pt}
The PTM method is very robust against thermal displacements and strain, and in high-temperature simulations it provides a significant improvement over existing identification methods.  We demonstrate this here with a simulated $\text{Cu}_3\text{Pt}$ alloy sample containing 2.8 million atoms, with periodic boundary conditions.  The sample initially contains 30 grains, with randomly selected centres and orientations.  The grain volumes have been constructed by computing the Voronoi cells of the grain centres.  The atoms within the grains are initially ordered with a perfect $\text{L1}_{2}$ structure.  This structure, which is the stable low-temperature phase of this system, can be considered as an FCC structure with three Cu atoms and one Pt atom in the unit cell.  It should therefore be identified as FCC by the PTM algorithm.

\begin{figure}[tbp]
\centering
\textbf{Polyhedral Template Matching}\\
\vspace{2mm}
\begin{minipage}{.32\textwidth}
  \centering
  \includegraphics[width=0.9\textwidth]{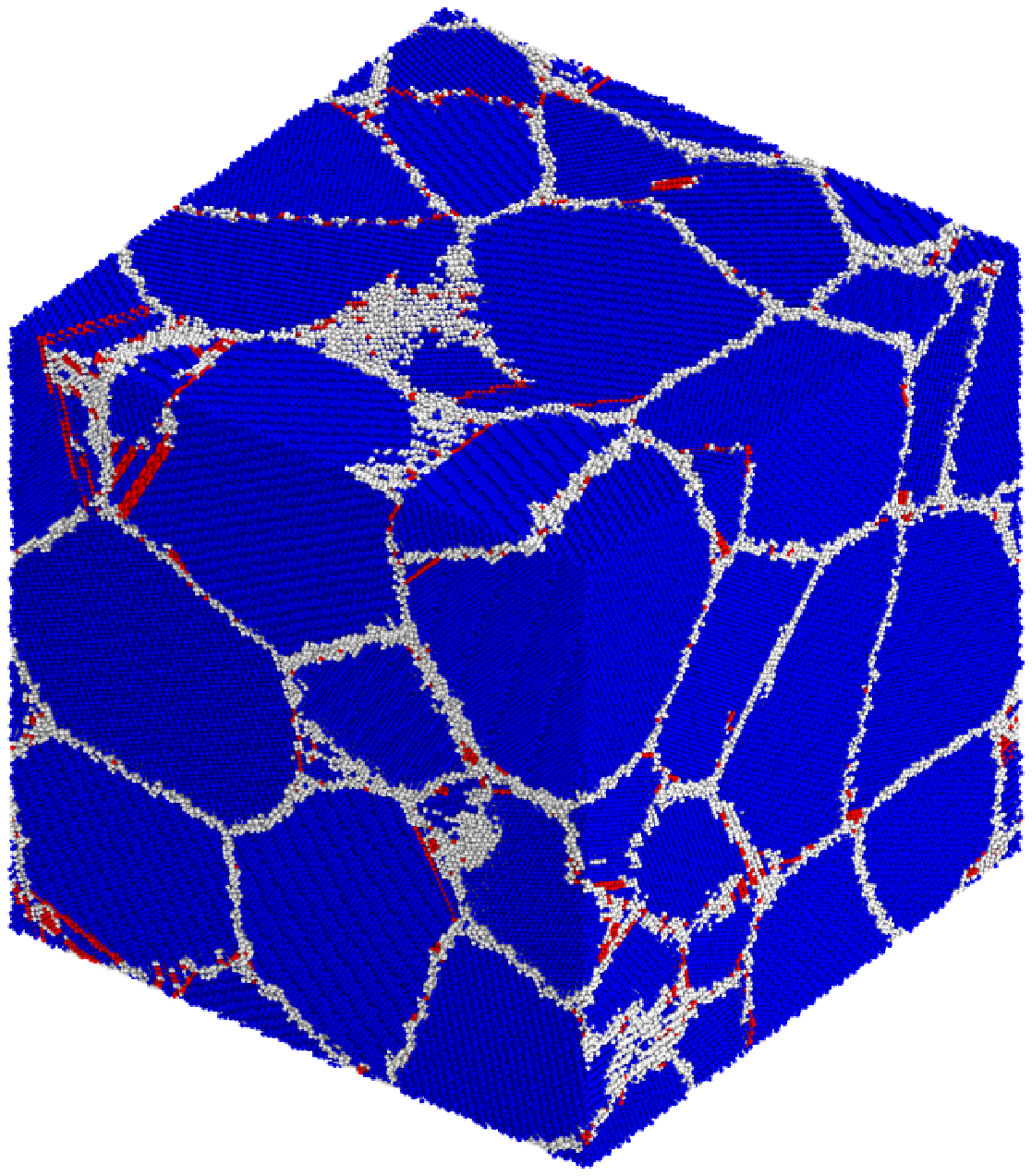}
\end{minipage}
\begin{minipage}{.32\textwidth}
  \centering
  \includegraphics[width=0.9\textwidth]{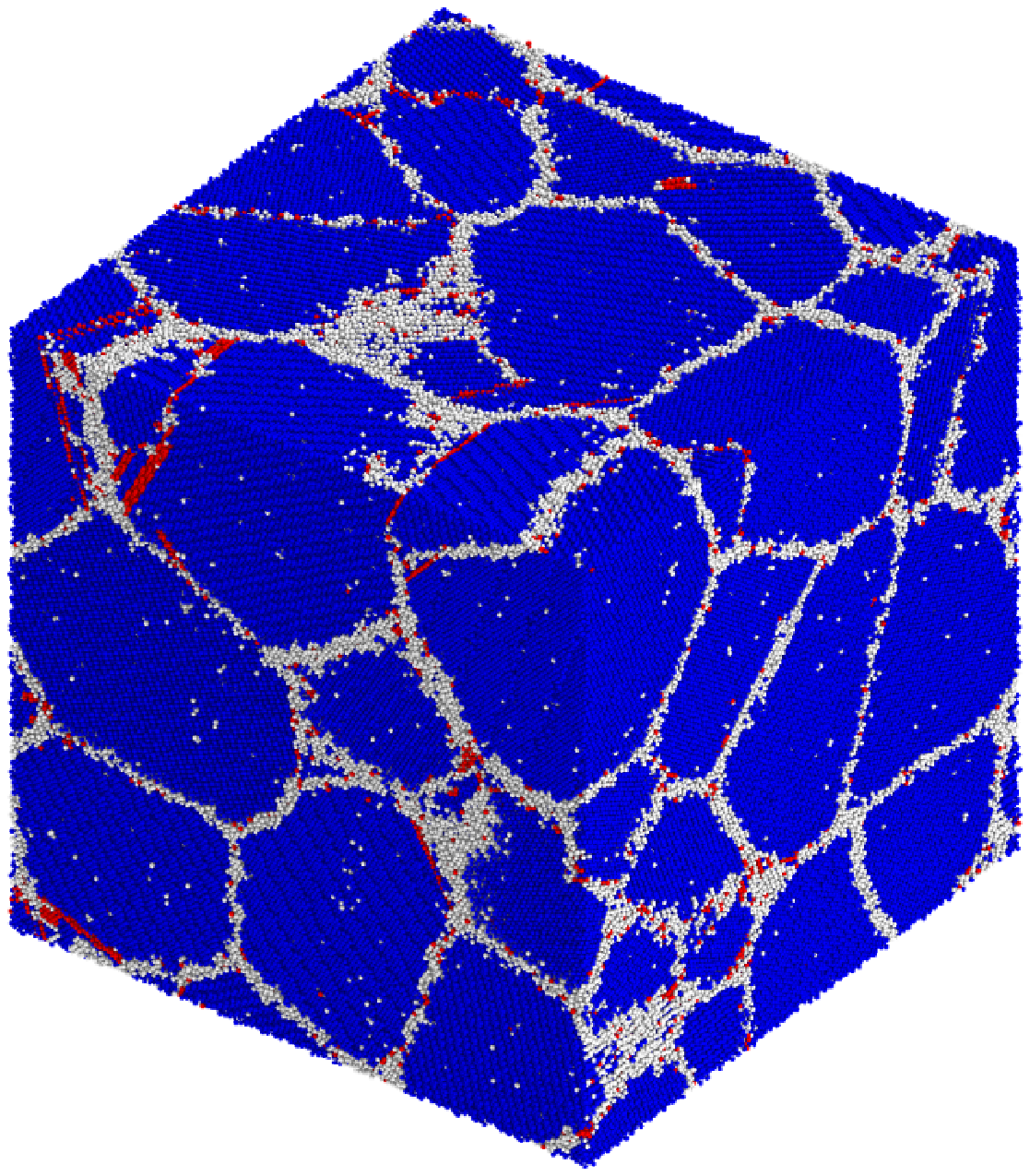}
\end{minipage}
\begin{minipage}{.32\textwidth}
  \centering
  \includegraphics[width=0.9\textwidth]{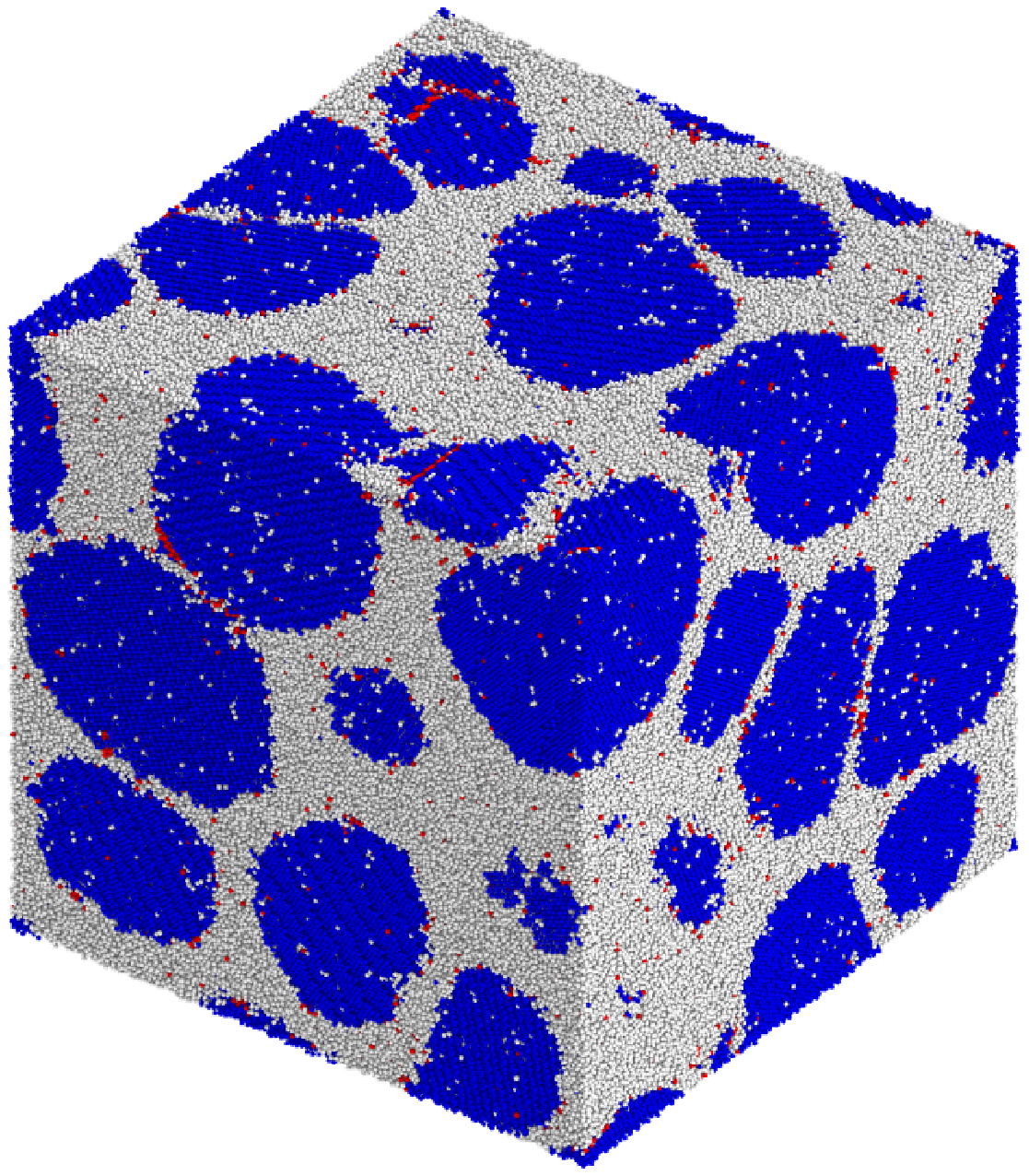}
\end{minipage}
\vspace{2mm}
\textbf{Adaptive Common Neighbour Analysis}\\
\vspace{2mm}
\begin{minipage}{.32\textwidth}
  \centering
  \includegraphics[width=0.9\textwidth]{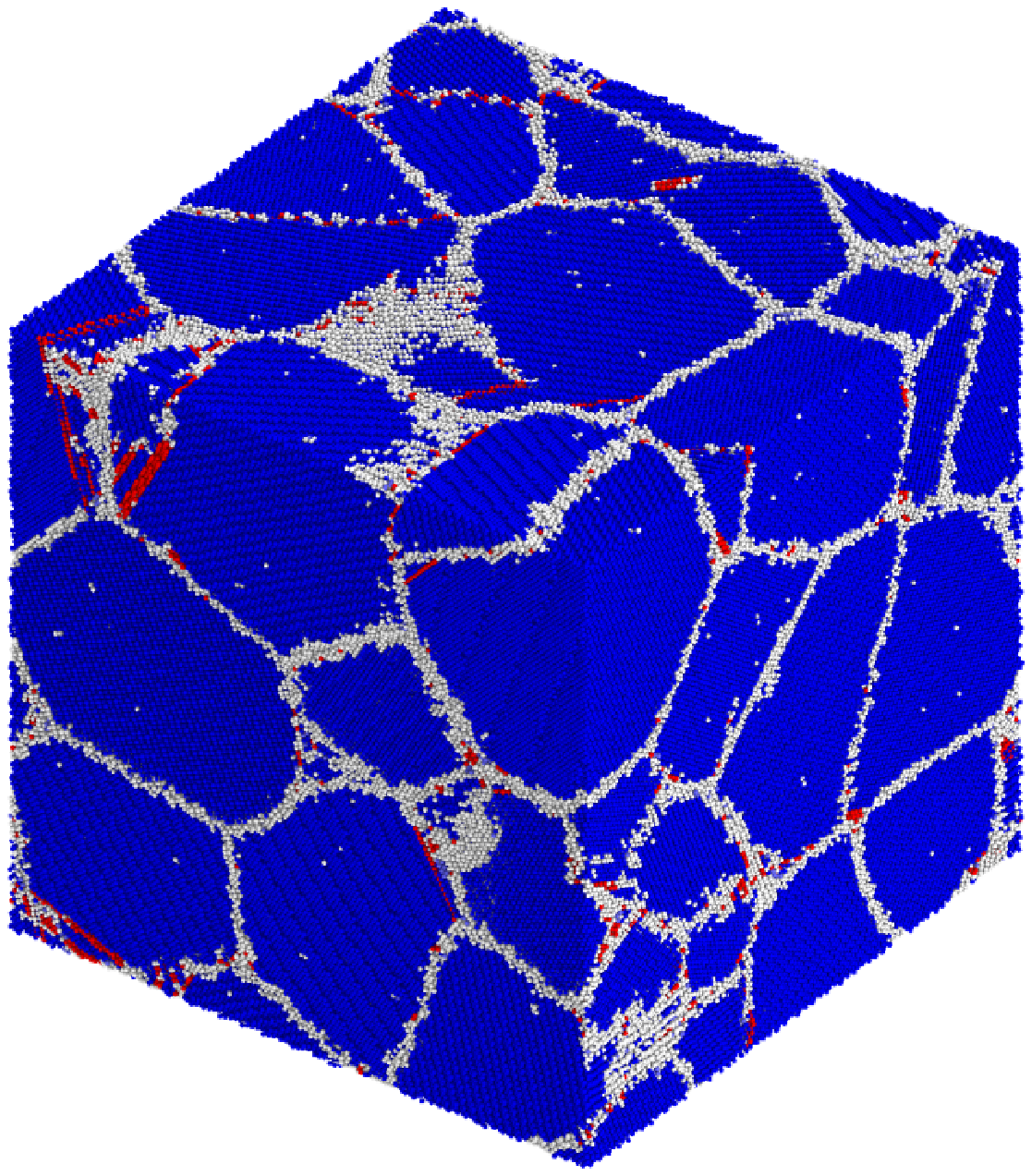}
  \textbf{500K}
\end{minipage}
\begin{minipage}{.32\textwidth}
  \centering
  \includegraphics[width=0.9\textwidth]{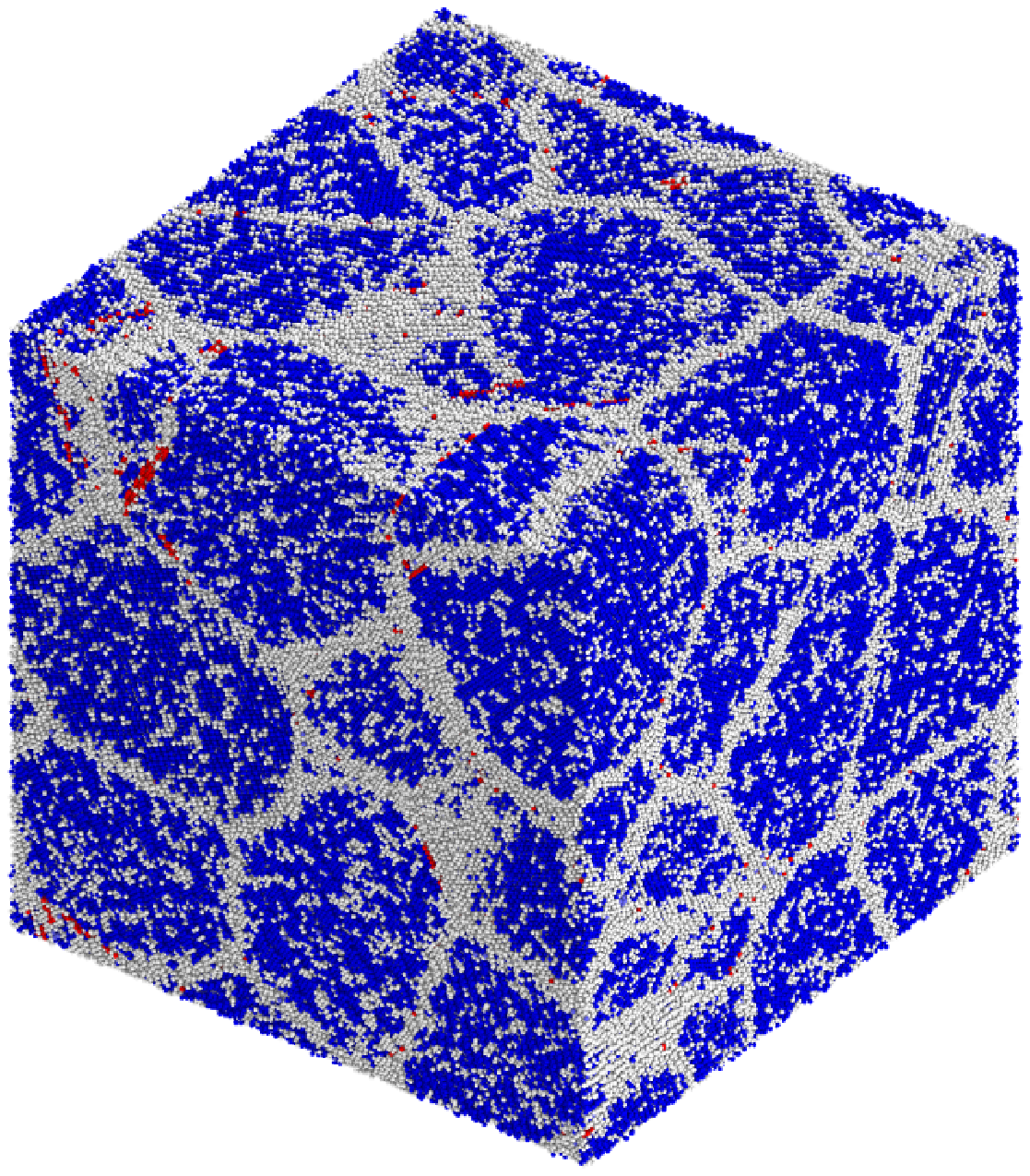}
  \textbf{1100K}
\end{minipage}
\begin{minipage}{.32\textwidth}
  \centering
  \includegraphics[width=0.9\textwidth]{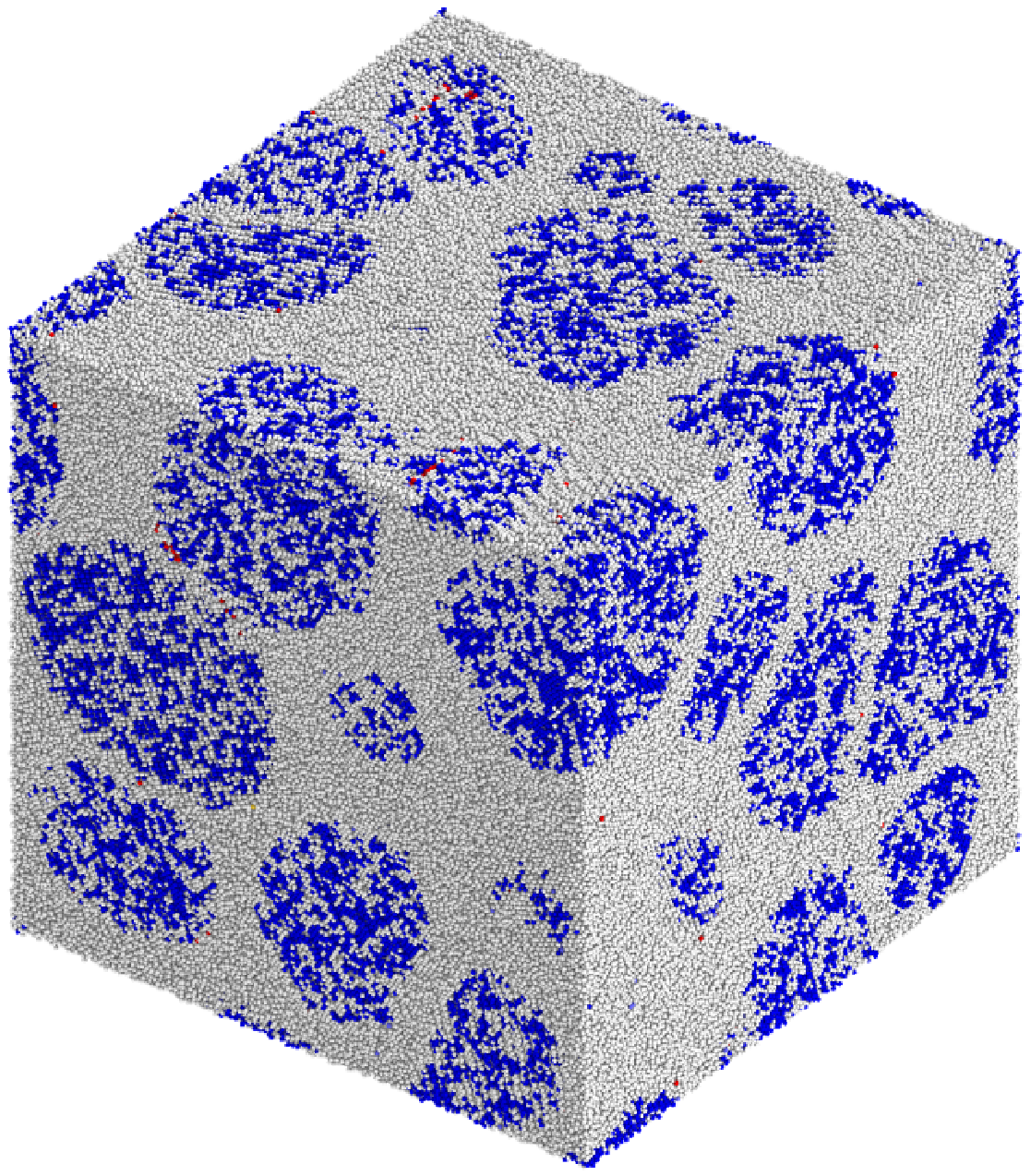}
  \textbf{1300K}
\end{minipage}
\\
\vspace{5mm}
\textbf{Legend:}\hspace{5mm}
\includegraphics[height=10pt]{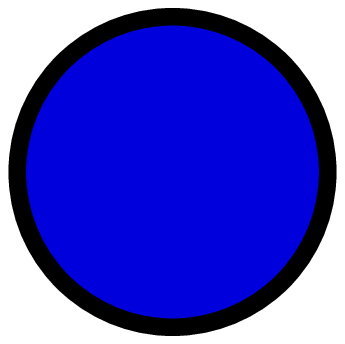} \textbf{FCC}
\hspace{4mm}
\includegraphics[height=10pt]{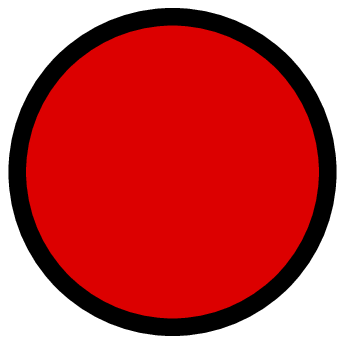} \textbf{HCP}
\hspace{4mm}
\includegraphics[height=10pt]{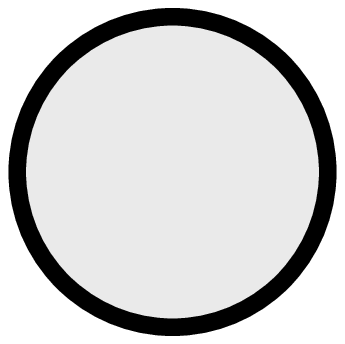} \textbf{Disordered}
\caption{Comparison of PTM with ACNA at three different temperatures on $\text{Cu}_3\text{Pt}$ systems containing 2.8 million atoms.  At 500K PTM offers little improvement over ACNA.  With increasing temperature the robustness of PTM is more evident.  At 1300K the system is in the process of melting from the grain boundaries.  Here, PTM finds twice the number of FCC atoms as ACNA.  The $\text{RMSD}_{\text{max}}$ values used for PTM are 0.11, 0.15 and 0.17 respectively.  The choice of RMSD values is motivated by the histograms in Figure~\ref{fig:cu3pt_fcc_rmsd_histograms}, but is not critical; using 0.17 for all figures would result in some grain boundary atoms identified as HCP.  The SC, ICO and BCC classifications are shown as disordered atoms.
The ACNA analyses shown here as well as all renders were performed with OVITO~\cite{Stukowski:2010ky}.
}
\label{fig:renders_comparison}
\end{figure}

The sample has first been quenched using the FIRE minimization method~\cite{Bitzek:2006bw}, and subsequently annealed at 1100K for 820 picoseconds using molecular dynamics with a 5 fs time step.  The interatomic potential is the Effective Medium Theory potential~\cite{Jacobsen:1996kb} and the temperature is controlled with a Berendsen thermostat~\cite{Berendsen:1984fm}.  After annealing, the sample has been separately heated to 1300K and cooled to 900K, 700K and 500K.
Figure~\ref{fig:renders_comparison} compares the performance of ACNA and PTM on the $\text{Cu}_3\text{Pt}$ systems at 500K and 1100K (both below the melting point temperature) and 1300K (near the melting point temperature).  Despite the system at 1300K being in the process of melting from the grain boundaries, PTM correctly identifies the majority of FCC structures in the non-melted volumes.  Topological ordering of neighbours improves detection of bulk crystallinity at high temperatures.
At 1300K, 9.6\% of the atoms identified as FCC with topological ordering were identified as disordered with distance ordering.

\begin{figure}[htbp]
\centering
\begin{minipage}[b]{.45\textwidth}
  \centering
  \includegraphics[width=1\linewidth]{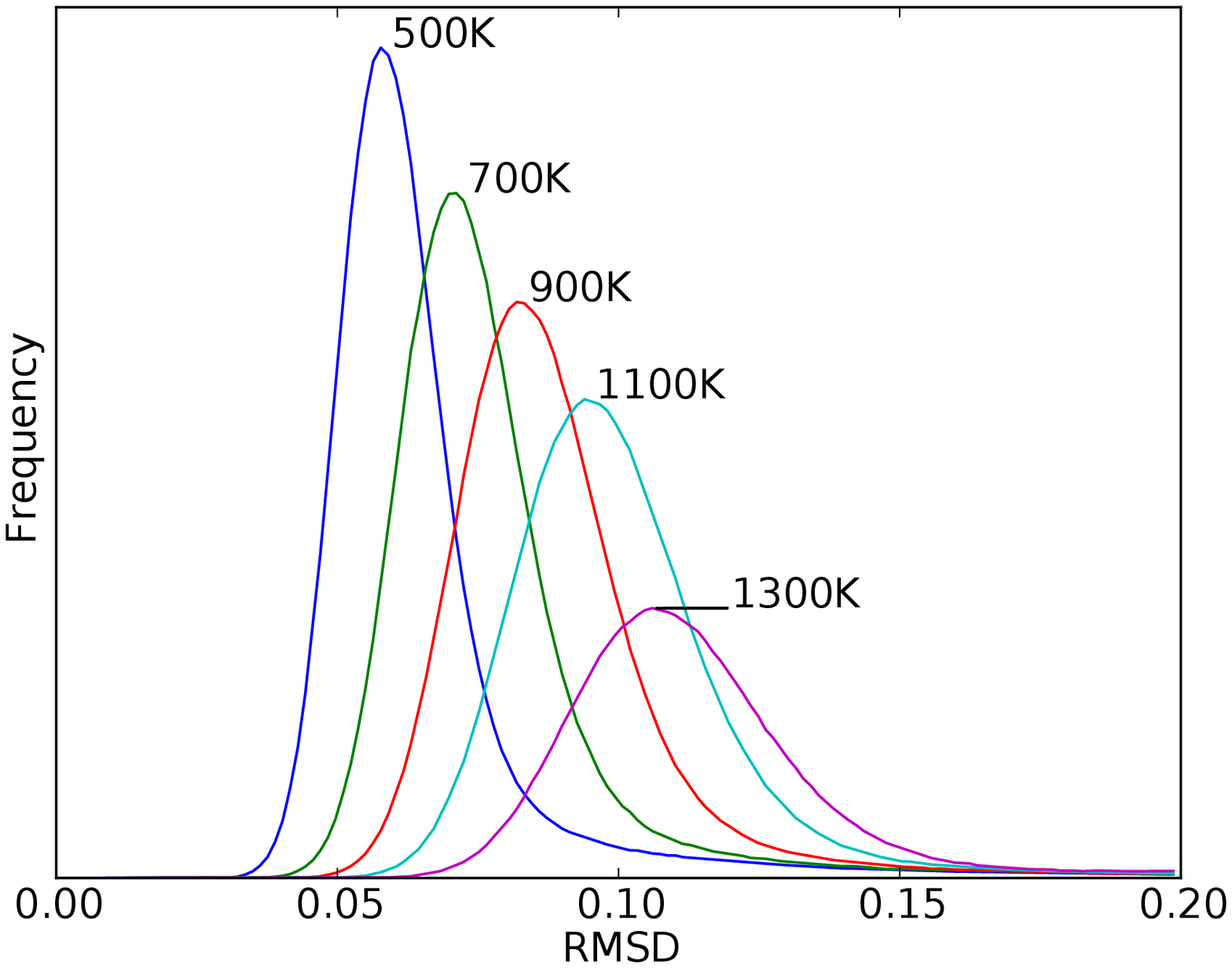}
\end{minipage}
\hspace{5mm}
\begin{minipage}[b]{.45\textwidth}
  \centering
  \includegraphics[width=1\linewidth]{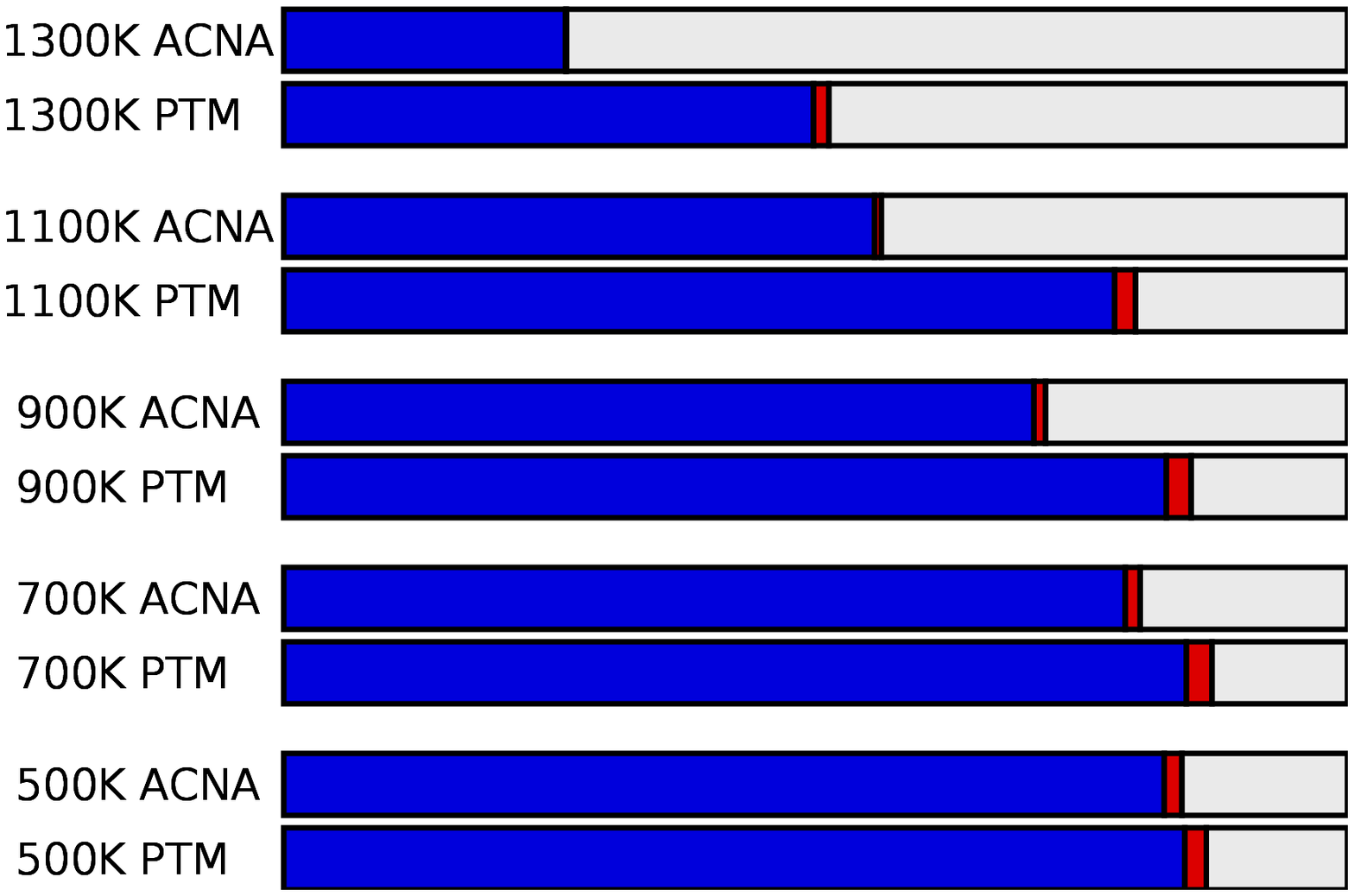}
  \\\vspace{2mm}
%\textbf{Legend:}\hspace{5mm}
\includegraphics[height=10pt]{render_legend_blue.eps} \textbf{FCC}
\hspace{4mm}
\includegraphics[height=10pt]{render_legend_red.eps} \textbf{HCP}
\hspace{4mm}
\includegraphics[height=10pt]{render_legend_grey.eps} \textbf{Disordered}
\end{minipage}
\begin{minipage}[t]{.45\textwidth}
\caption{Histograms of RMSD values for the FCC atoms in a polycrystalline $\text{Cu}_3\text{Pt}$ system at different temperatures.  With increasing temperature the thermal displacements result in larger RMSD values.}
\label{fig:cu3pt_fcc_rmsd_histograms}
\end{minipage}
\hspace{5mm}
\begin{minipage}[t]{.45\textwidth}
  \caption{Relative proportions of FCC, HCP and disordered structures found with ACNA and PTM at different temperatures. The SC, ICO and BCC classifications have been counted as disordered.}
  \label{fig:method_bar_chart}
\end{minipage}
\end{figure}

Figure~\ref{fig:cu3pt_fcc_rmsd_histograms} shows the effect of temperature on the RMSD distribution of the structures identified as FCC.  Higher temperatures lead to more distorted structures.  By choosing good values for $\text{RMSD}_{\text{max}}$, spurious identifications (in this case SC, ICO and BCC) are significantly reduced.  Figure~\ref{fig:method_bar_chart} shows the proportion of each structure type identified for the systems shown in Figure~\ref{fig:renders_comparison}.
%The full data as well as the $\text{RMSD}_{\text{max}}$ values used are given in Table~\ref{table:comparison}. (this table has now been commented out)

%\FloatBarrier
\section{Alloy Structures}
\label{sec:alloys}
\begin{figure}[tbp]
\centering
\begin{minipage}{.19\textwidth}
  \centering
  \includegraphics[width=0.9\textwidth]{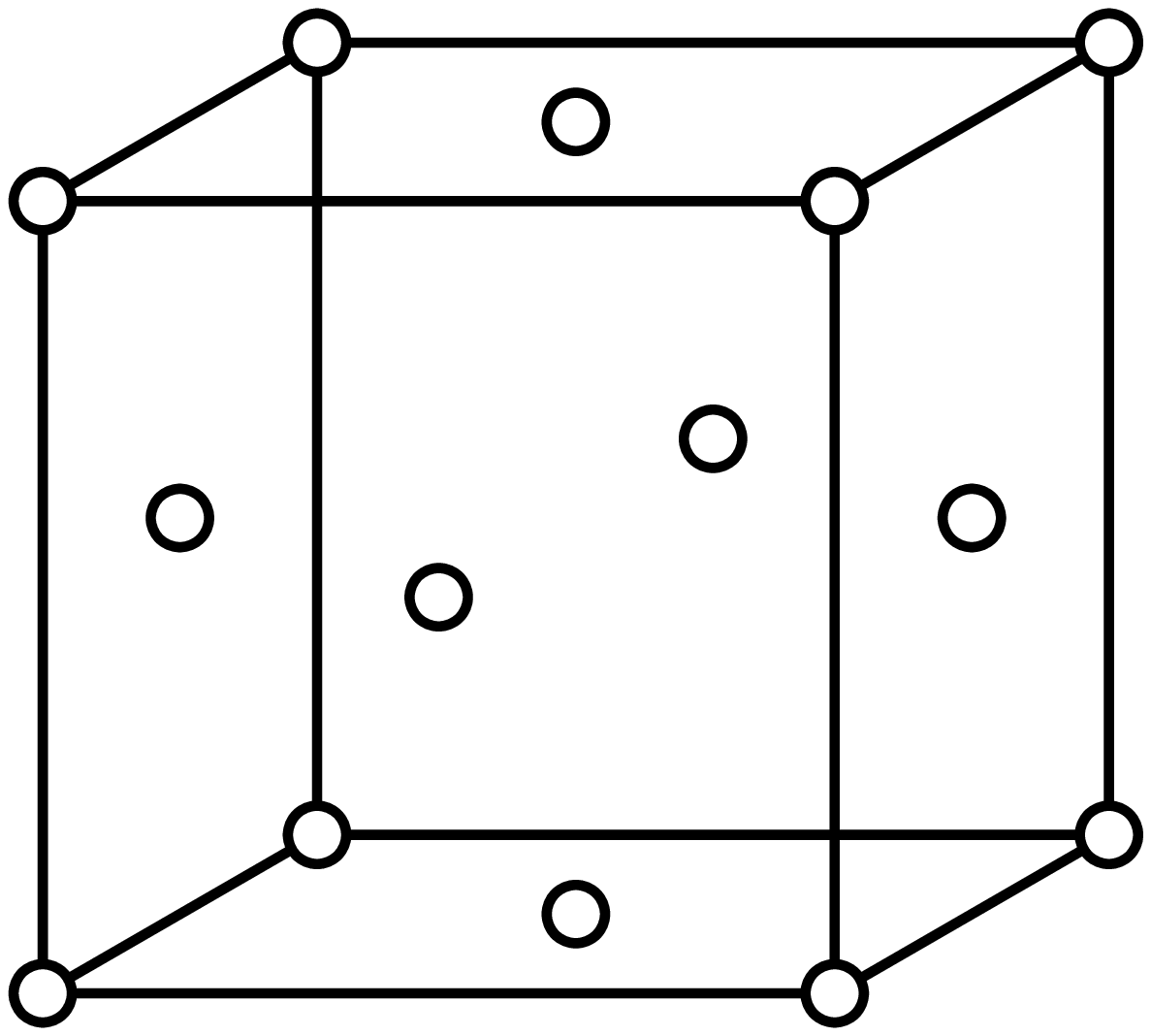}\\
  \sublabel{a} A1 (Cu)
\end{minipage}
\begin{minipage}{.19\textwidth}
  \centering
  \includegraphics[width=0.9\textwidth]{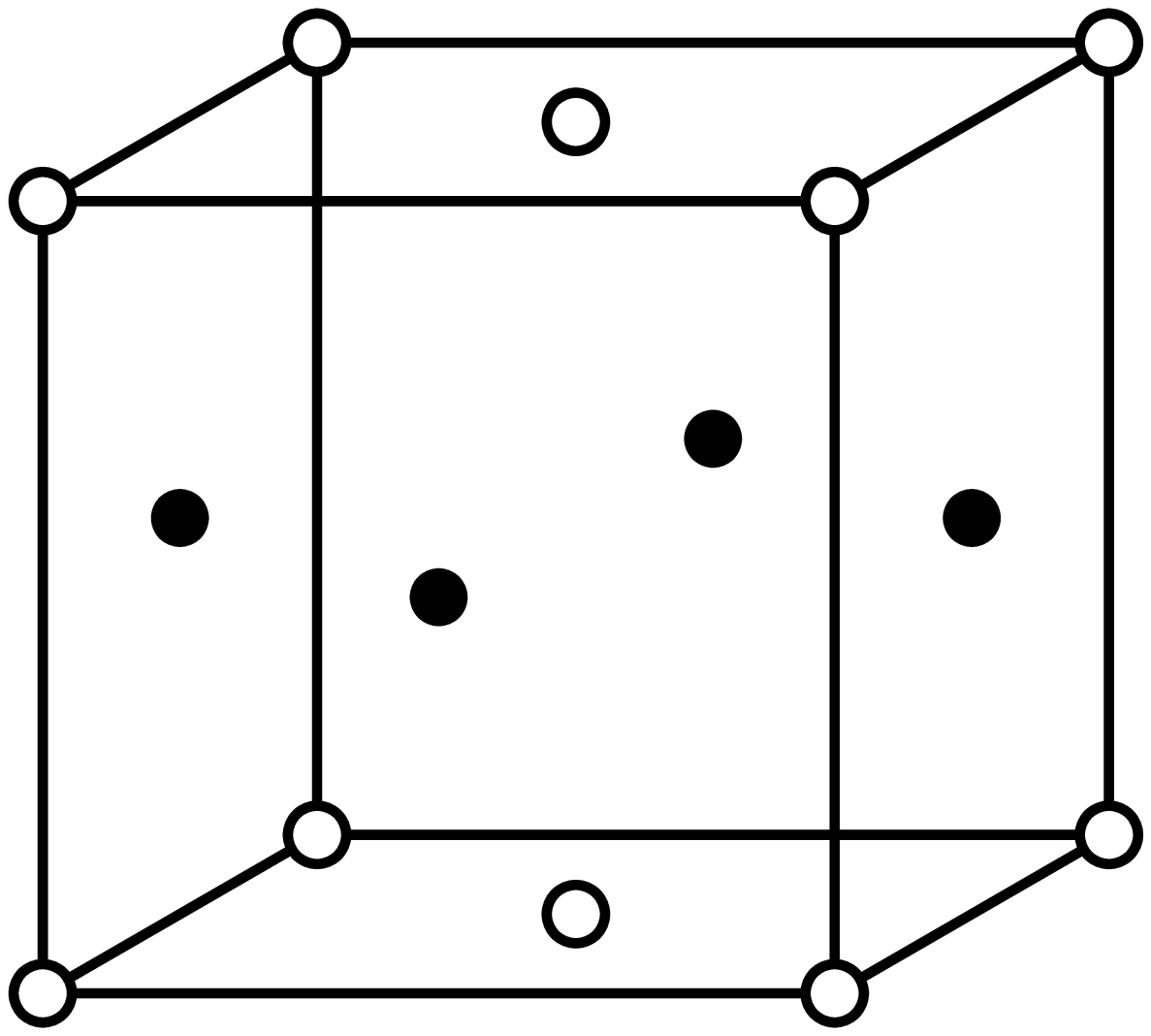}\\
  \sublabel{b} $\text{L}1_0$ $(\text{Cu}\text{Au})$
\end{minipage}
\begin{minipage}{.19\textwidth}
  \centering
  \includegraphics[width=0.9\textwidth]{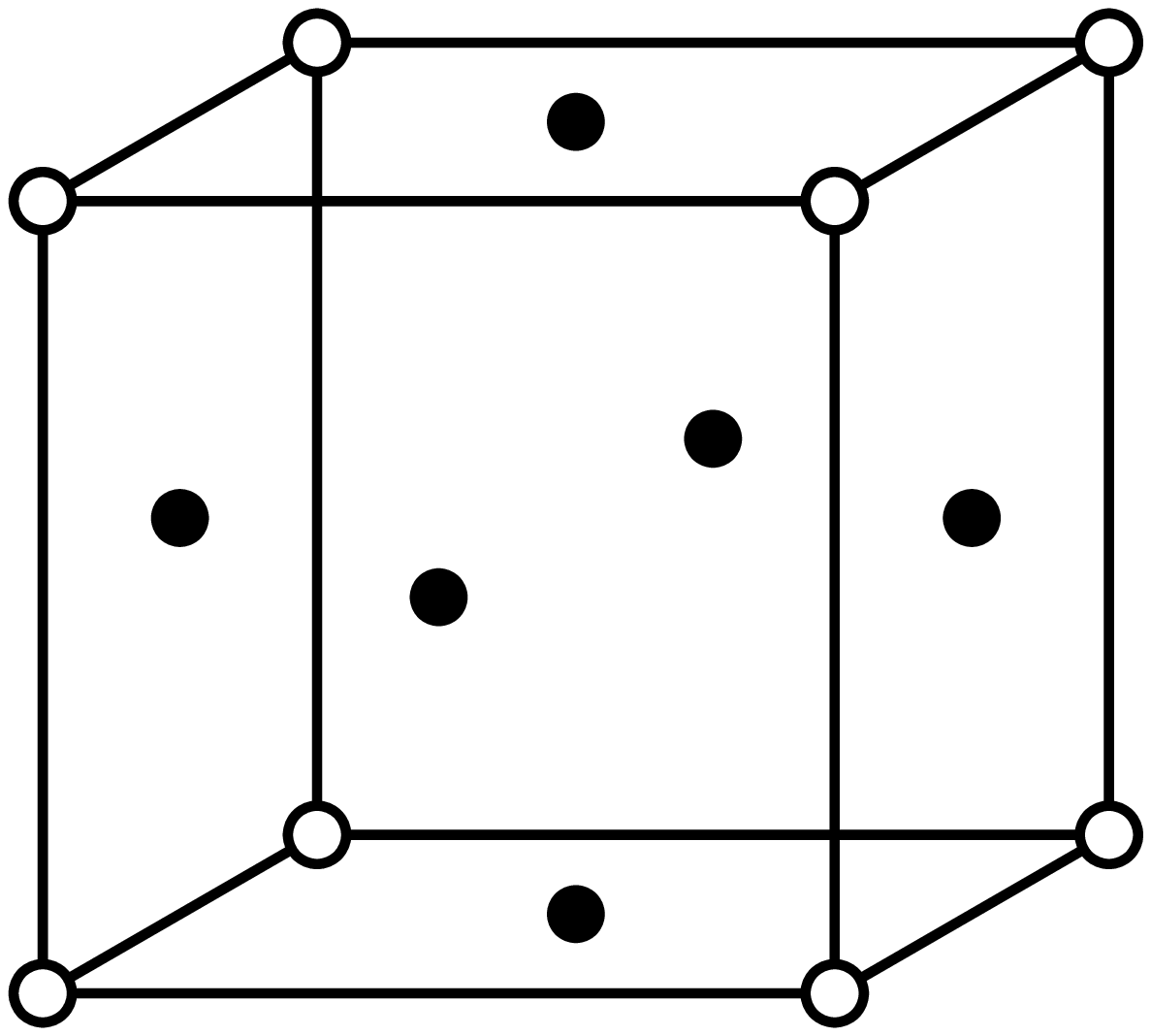}\\
  \sublabel{c} $\text{L}1_2$ $(\text{Cu}_3\text{Au})$
\end{minipage}
\begin{minipage}{.19\textwidth}
  \centering
  \includegraphics[width=0.9\textwidth]{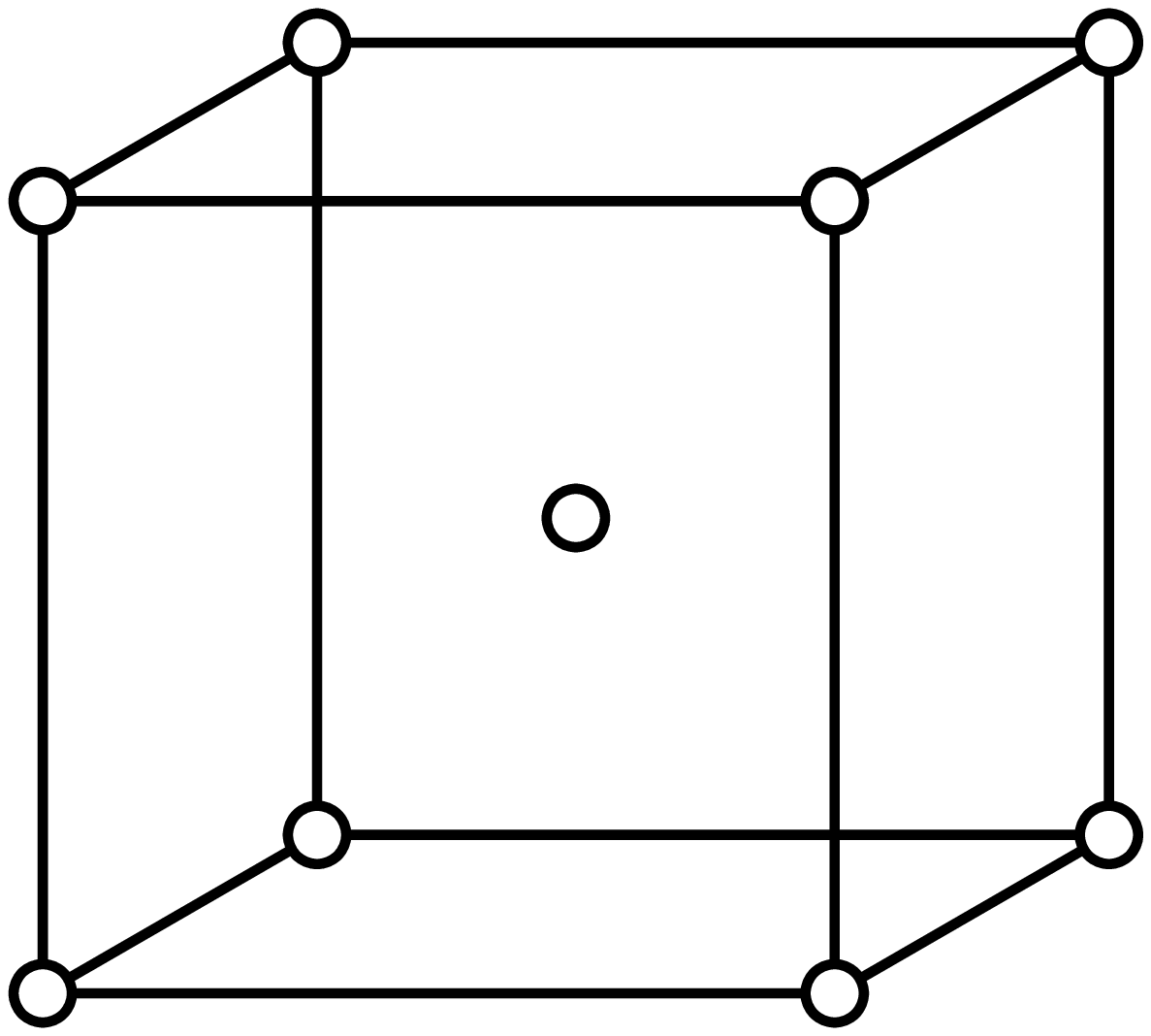}\\
  \sublabel{d} A2 (W)
\end{minipage}
\begin{minipage}{.19\textwidth}
  \centering
  \includegraphics[width=0.9\textwidth]{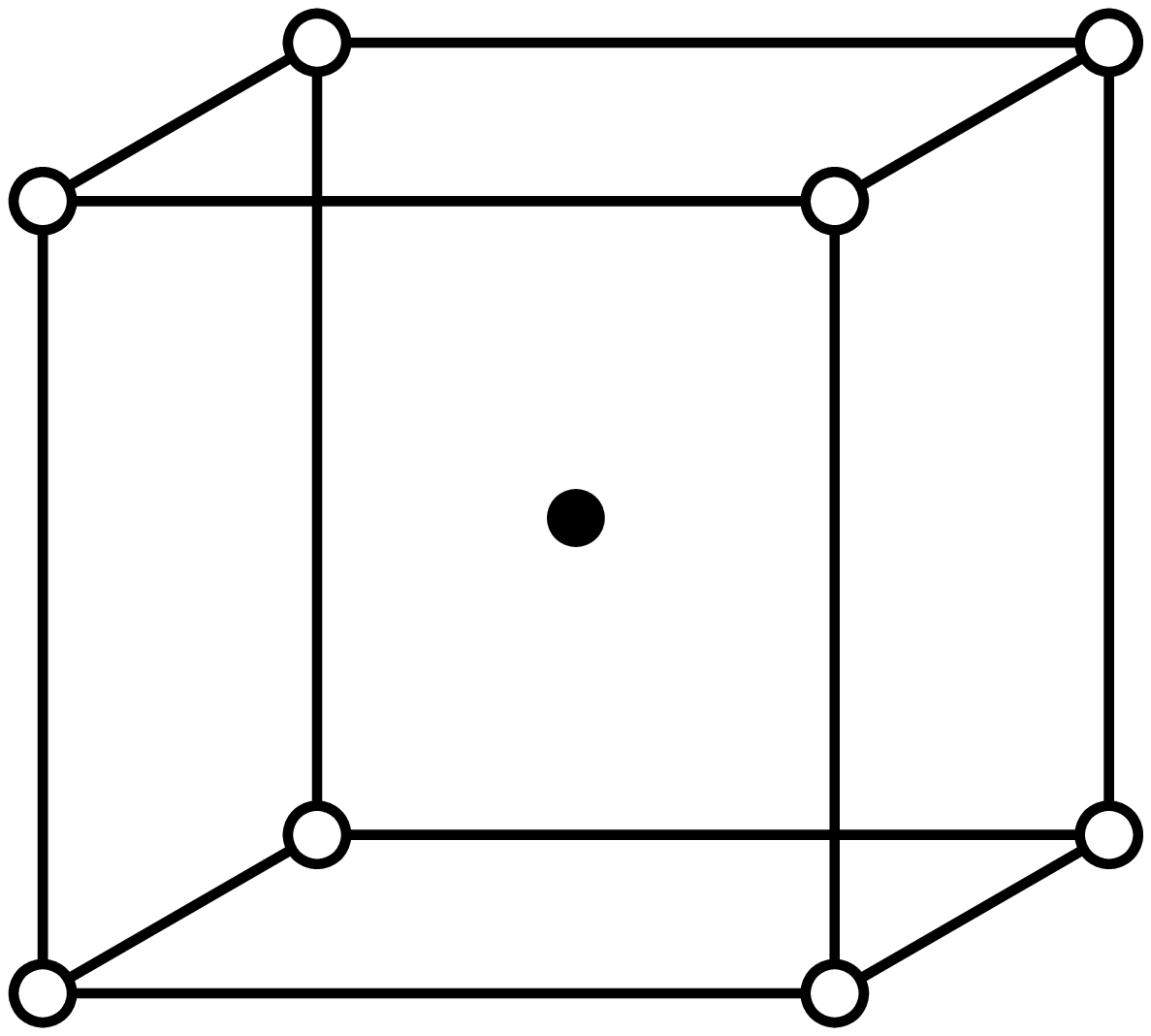}\\
  \sublabel{e} B2 (CsCl)
\end{minipage}
\caption{Lattice structures of the three FCC lattice types (a)-(c), and the two BCC lattice types (d)-(e).}
\label{fig:alloy_structures_FCC_BCC}
\end{figure}

Since PTM finds the optimal point-to-point correspondences between a set of points and an ideal reference template, we can easily identify alloy structures.  A good description of the possible lattice structures of FCC and BCC alloys is given in~\cite{laughlin1988long}.  Some figures are recreated in Figure~\ref{fig:alloy_structures_FCC_BCC}.  Here, we will only consider binary alloys, though multi-element alloys would be a simple extension.

\begin{figure}[bp]
\centering
\begin{minipage}{.19\textwidth}
  \centering
  \includegraphics[width=0.9\textwidth]{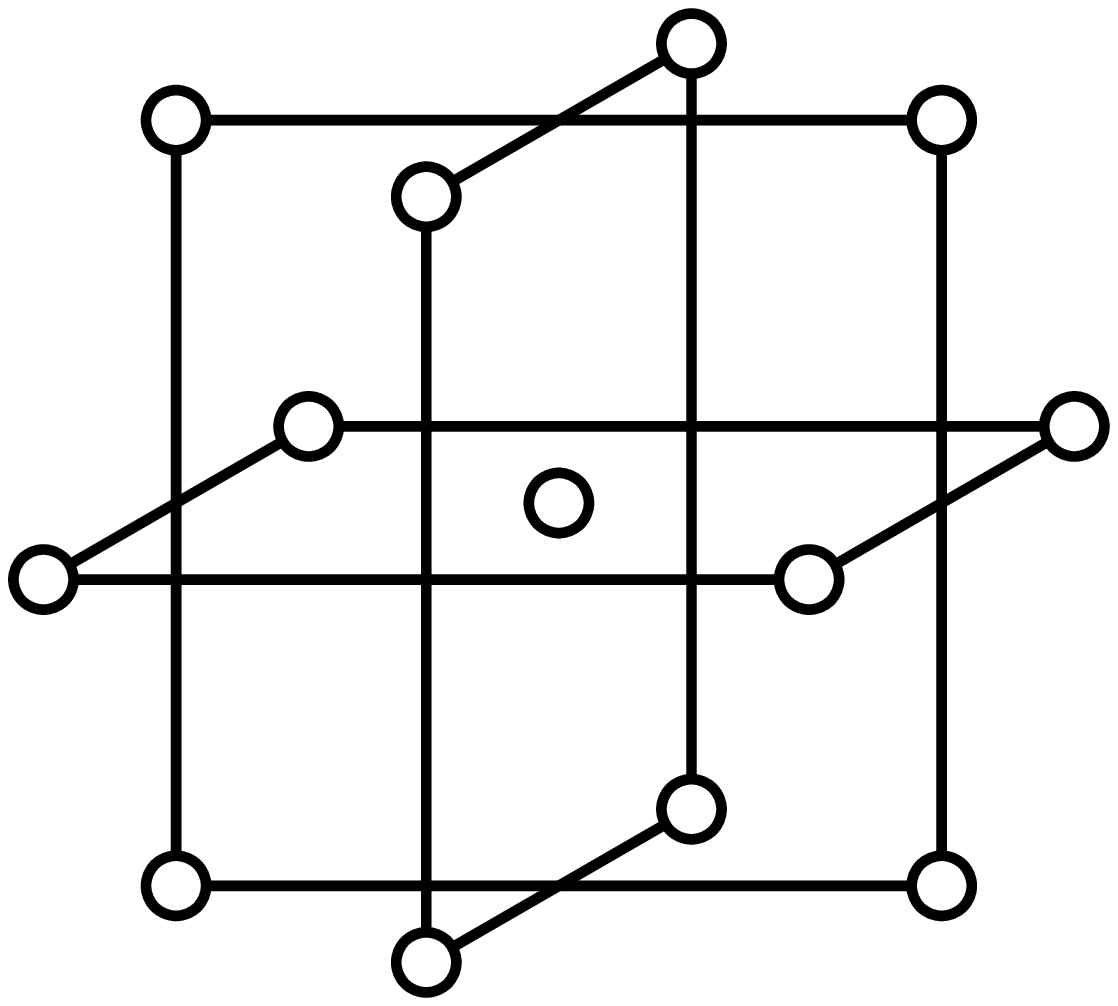}\\
  \sublabel{a} A1 Cu-type
\end{minipage}
\begin{minipage}{.19\textwidth}
  \centering
  \includegraphics[width=0.9\textwidth]{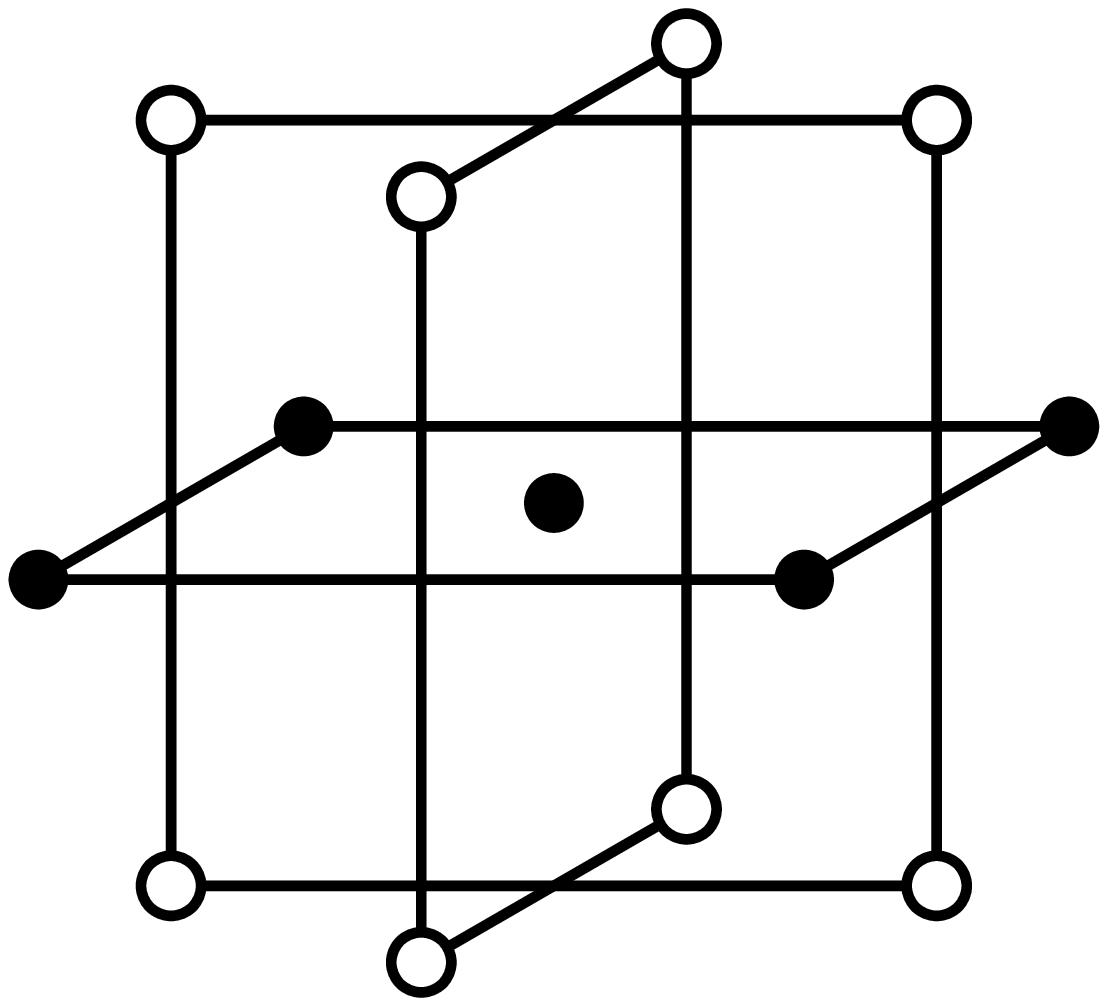}\\
  \sublabel{b} $\text{L}1_0$ Cu-type
\end{minipage}
\begin{minipage}{.19\textwidth}
  \centering
  \includegraphics[width=0.9\textwidth]{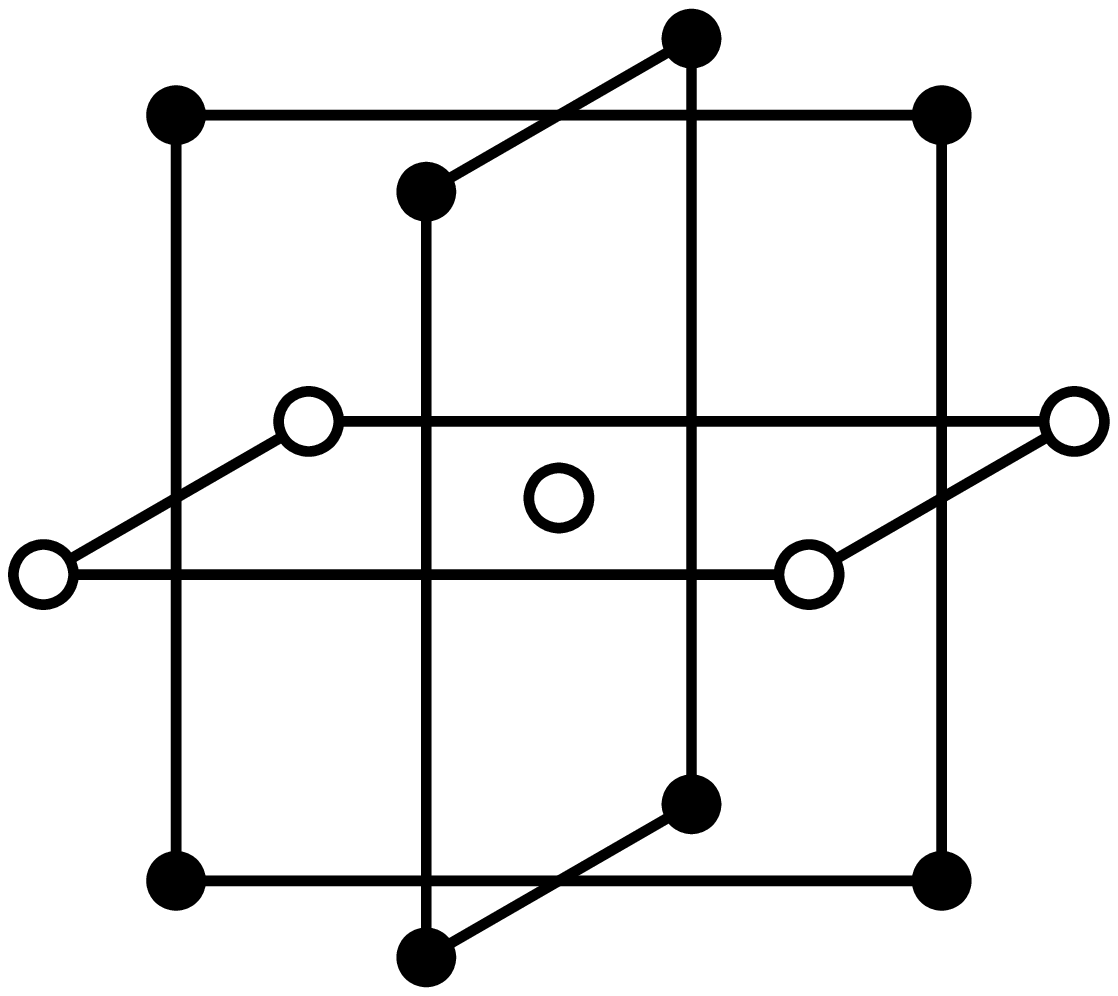}\\
  \sublabel{c} $\text{L}1_0$ Au-type
\end{minipage}
\begin{minipage}{.19\textwidth}
  \centering
  \includegraphics[width=0.9\textwidth]{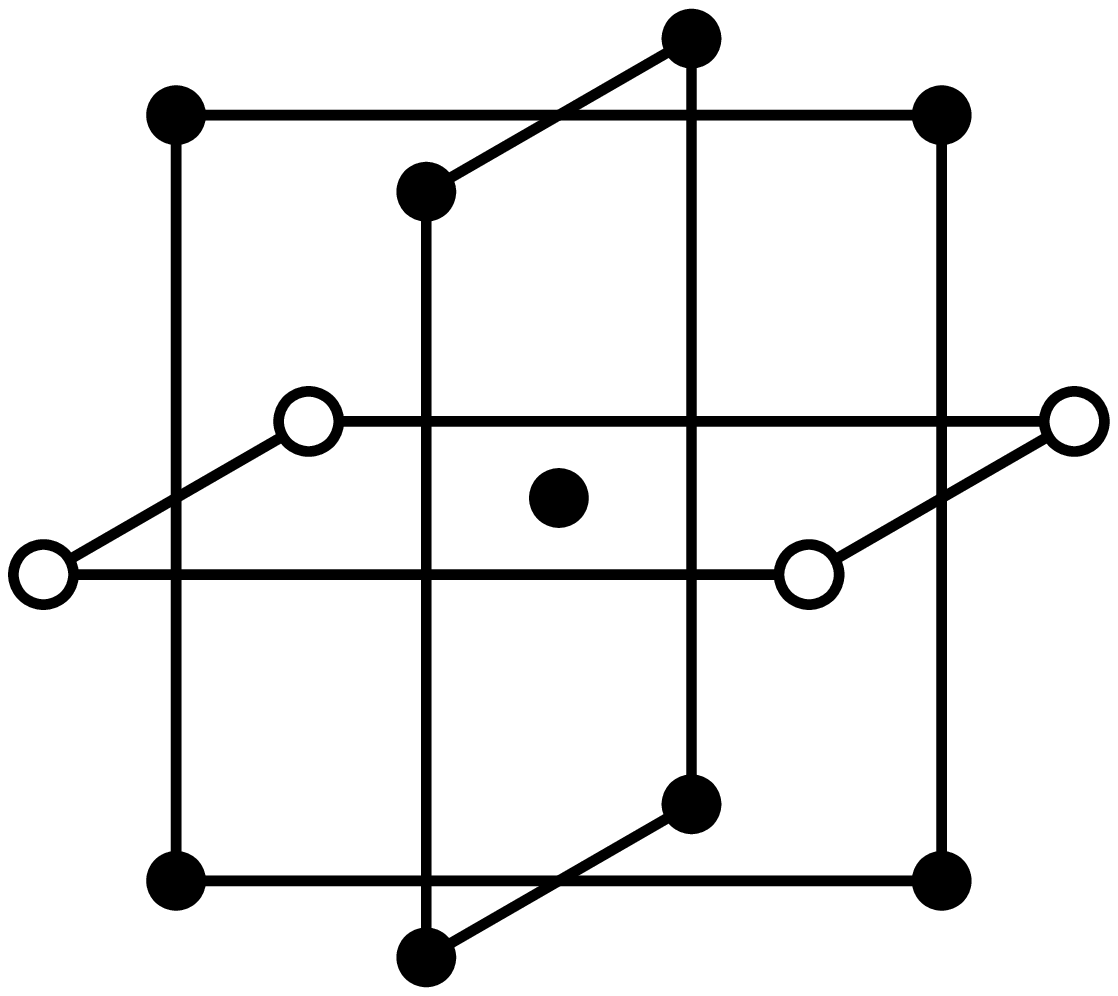}\\
  \sublabel{d} $\text{L}1_2$ Cu-type
\end{minipage}
\begin{minipage}{.19\textwidth}
  \centering
  \includegraphics[width=0.9\textwidth]{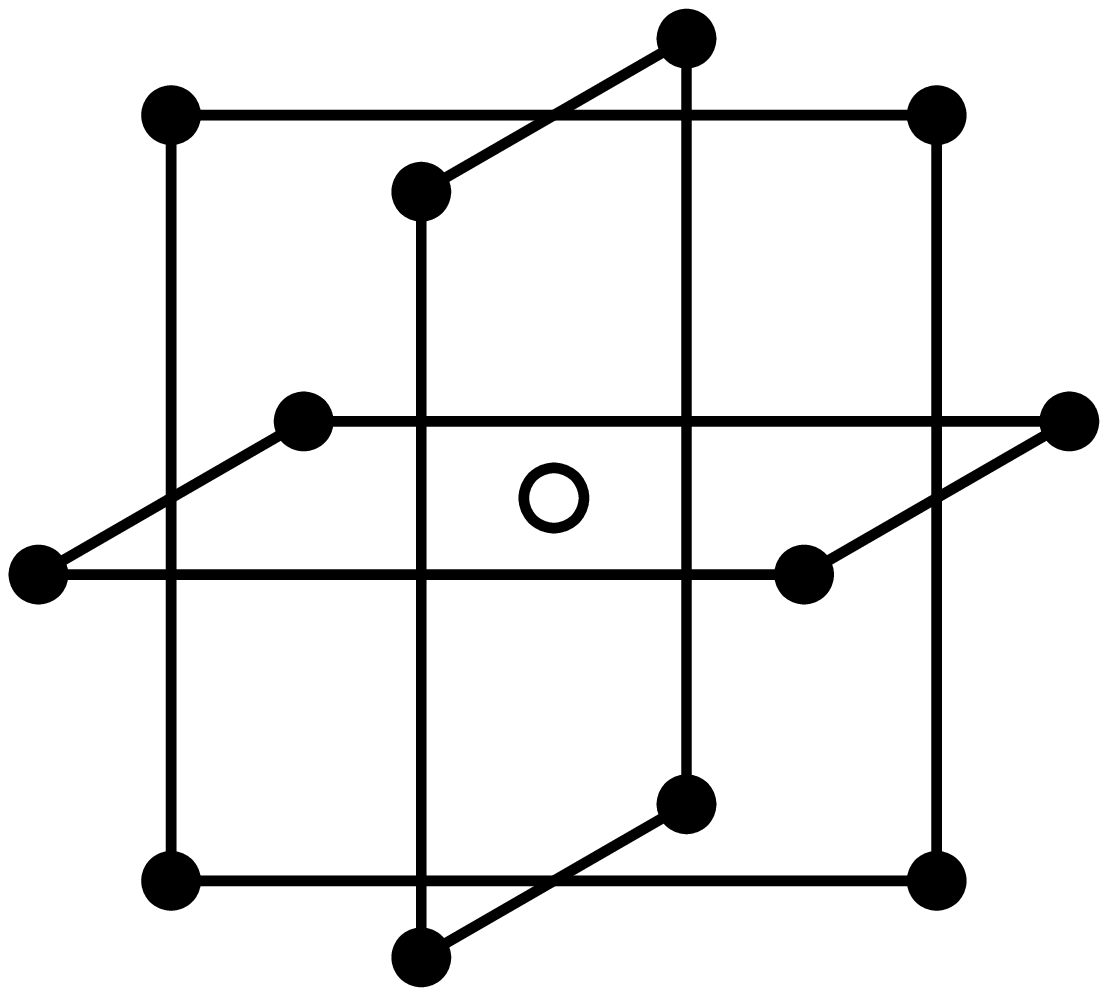}\\
  \sublabel{e} $\text{L}1_2$ Au-type
\end{minipage}
\caption{Alloy lattice structures of a central atom and the first shell of nearest neighbours in a FCC lattice, for different central atom types.  Different alloy types can be determined by counting the atom types of coplanar neighbours.}
\label{fig:alloy_structures_central}
\end{figure}

Identification of the A1 and A2 structures is trivial; all nearest neighbours must have the same atom type as the central atom.  The B2 structure is equally simple to identify; the types of all the atoms in the first shell of neighbours must be the opposite of the central atom, and all the atoms in the second shell must have the same type as the central atom.  For the $\text{L}1_0$ and $\text{L}1_2$ structures, it is instructive to view the structures from the central atom.  This is shown in Figure~\ref{fig:alloy_structures_central}.

Using the point-to-point correspondences, the atom types of a distorted structure can be mapped onto the ideal FCC lattice structure.  The alloy type can then be determined by examining the types of coplanar neighbours.  Table~\ref{table:alloy_rules} summarizes the rules for determining FCC alloy structures.

\begin{table}[tbp]
\centering
\begin{tabular}{|c|c|l|}
\hline
Alloy structure		& Atom Type	& Neighbours\\
\hline
A1					& Cu		& 3 x 4 coplanar Cu-type (all Cu-type)\\
A1					& Au		& 3 x 4 coplanar Au-type (all Au-type)\\
$\text{L}1_0$		& Cu		& 2 x 4 coplanar Au-type, 4 coplanar Cu-type\\
$\text{L}1_0$		& Au		& 2 x 4 coplanar Cu-type, 4 coplanar Au-type\\
$\text{L}1_2$		& Cu		& 2 x 4 coplanar Cu-type, 4 coplanar Au-type\\
$\text{L}1_2$		& Au		& 3 x 4 coplanar Cu-type (all Cu-type)\\
\hline
\end{tabular}
\caption{Rules for determining FCC binary alloy types using the types of coplanar neighbours.
\label{table:alloy_rules}}
\end{table}

\begin{figure}[tbp]
\centering
\begin{minipage}{.45\textwidth}
  \centering
  \includegraphics[width=0.75\textwidth]{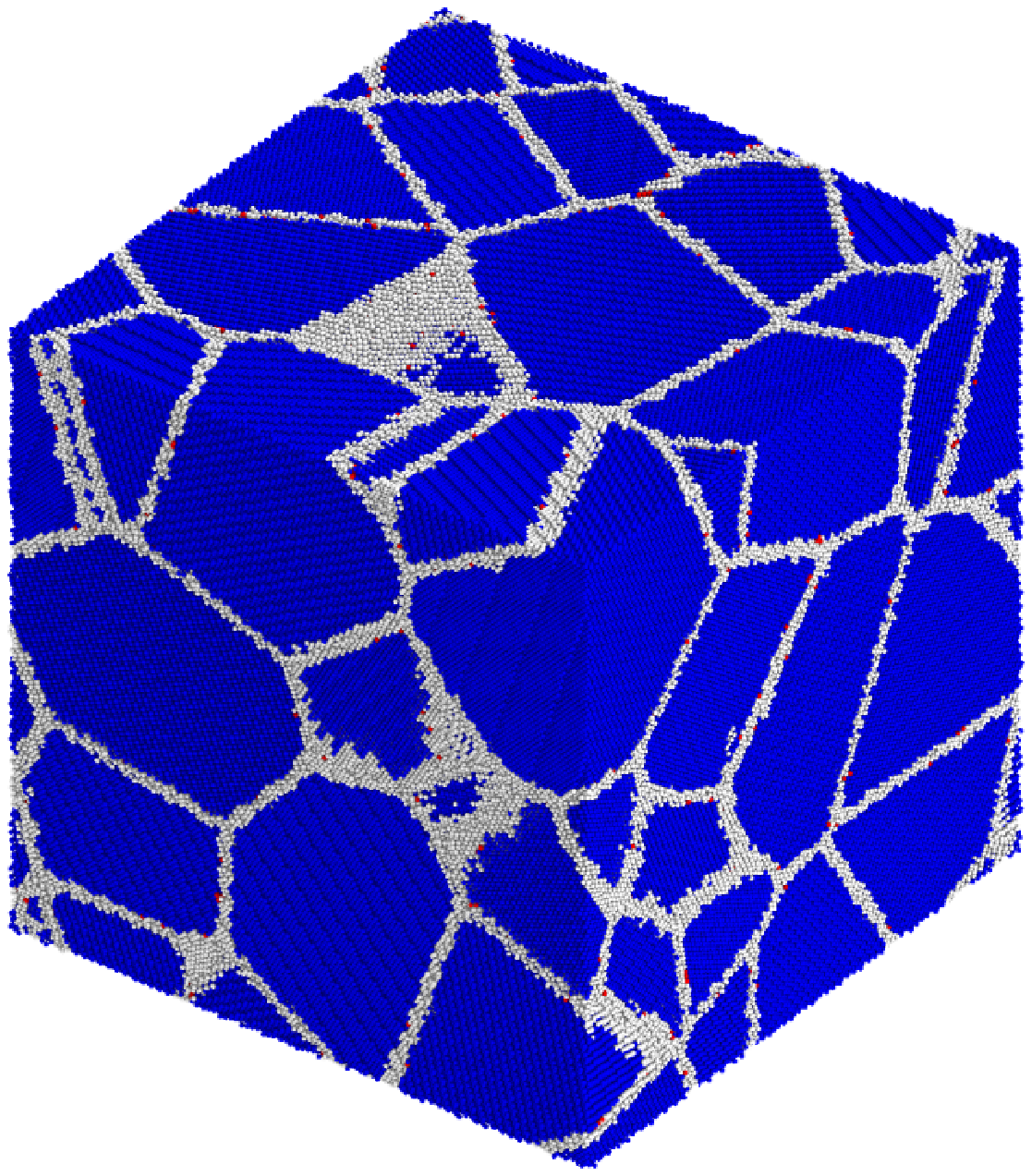}\\
  \textbf{Pre-annealing}
\end{minipage}
\hspace{5mm}
\begin{minipage}{.45\textwidth}
  \centering
  \centering
  \includegraphics[width=0.75\textwidth]{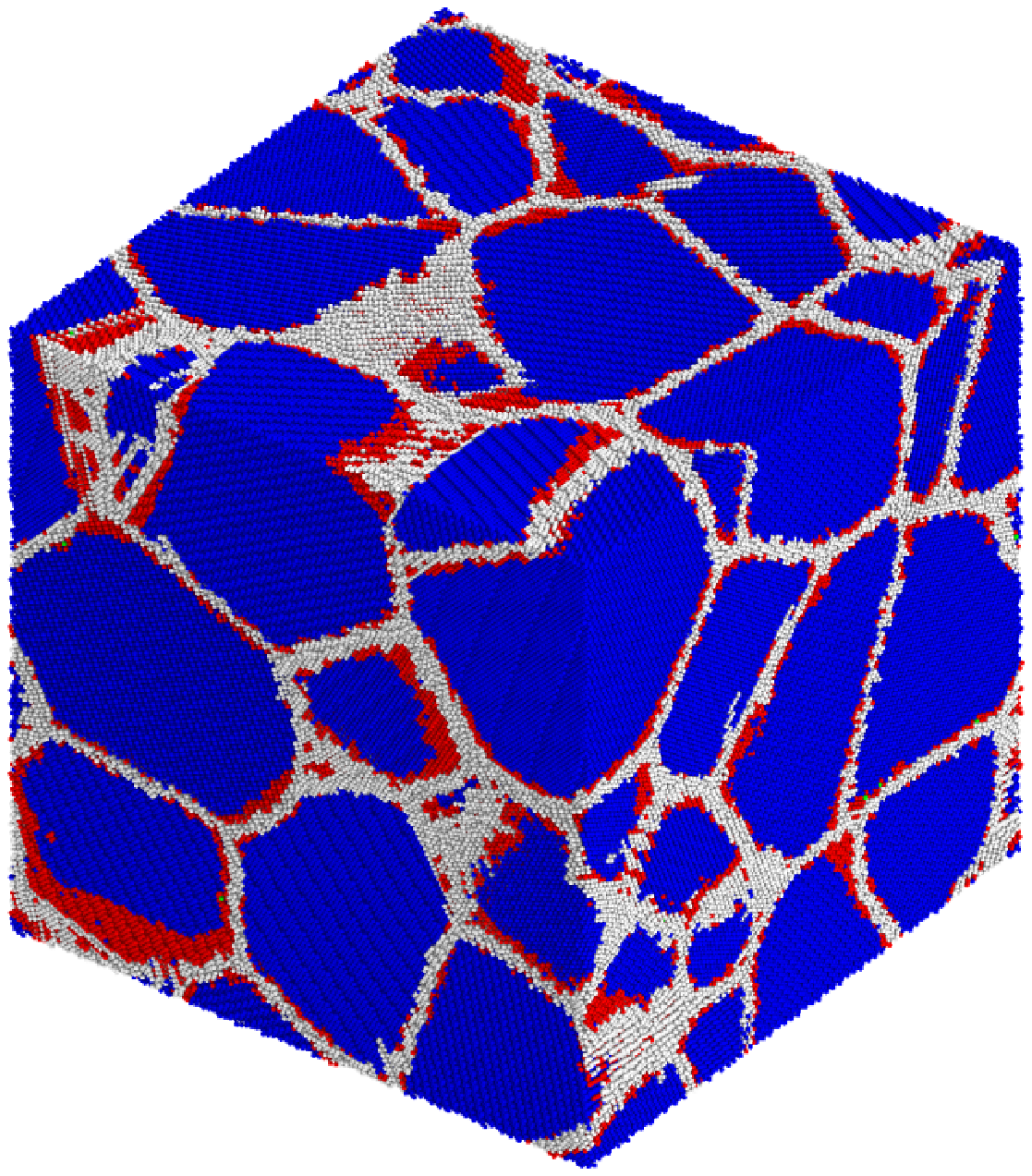}\\
  \textbf{Post-annealing}
\end{minipage}
\\
\vspace{5mm}
\textbf{Legend:}\hspace{5mm}
\includegraphics[height=10pt]{render_legend_blue.eps} \textbf{$\text{L}\mathbf{1_2}$ FCC}
\hspace{4mm}
\includegraphics[height=10pt]{render_legend_red.eps} \textbf{Disordered FCC}
\hspace{4mm}
\includegraphics[height=10pt]{render_legend_grey.eps} \textbf{Non-FCC}\\
\caption{Alloy structures in quenched polycrystalline $\text{Cu}_3\text{Pt}$ systems.  Before annealing (left), the grains have been initialized with perfect $\text{L}1_{2}$ ordering.  After annealing at 1100K for 820 picoseconds (right), the atoms which have passed through a grain boundary have disordered FCC alloy structure.  The cut-off used is $\text{RMSD}_{\text{max}} = 0.05$.}
\label{fig:renders_alloys}
\end{figure}

To illustrate the method described here, we have determined the alloy structures of the $\text{Cu}_3\text{Pt}$ system described in section~\ref{sec:results_cu3pt}, both before annealing and after annealing.  After the sample grains have been initialized by Voronoi cell construction, with perfect $\text{L}1_{2}$ ordering, the sample has been quenched.  The sample has also been quenched, post-annealing, to allow the alloy structures to be compared.  The alloy structures of each system are shown in Figure~\ref{fig:renders_alloys}.  It can be seen that grain boundary migration affects the alloy structure; despite the atoms still having FCC structure after passing through a grain boundary, the $\text{L}1_{2}$ ordering has been lost.  
Before annealing, less than 0.5\% of the FCC atoms were not identified as being in the $\text{L}1_{2}$ structure.  After annealing, this percentage had grown to approximately 12\%.

%\FloatBarrier
\section{Lattice Orientations}
\label{sec:orientations}
\begin{figure}[tp]
\centering
\includegraphics[width=0.4\textwidth]{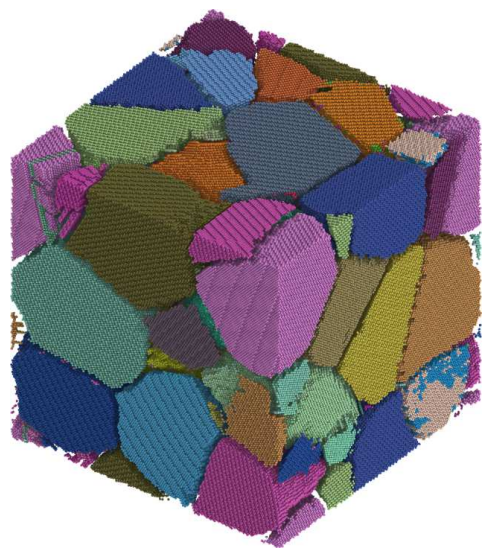}
\caption{Orientations of FCC atoms in a polycrystalline $\text{Cu}_3\text{Pt}$ system, with $\text{RMSD}_{\text{max}} = 0.05$.}
\label{fig:render_orientation}
\end{figure}

A consequence of PTM identifying structures by minimizing the RMSD (c.f. Equation~(\ref{eq:rmsd_scale_invariant})) is that the orientation is defined for each atom, at no extra computational cost.  Furthermore, due to the optimal point-to-point correspondence being determined, the orientation is robustly determined.  Figure~\ref{fig:render_orientation} shows the lattice orientation of the FCC atoms in the $\text{Cu}_3\text{Pt}$ system, post-annealing.  The non-FCC atoms have not been included in the render.
The colours used are obtained by rotation of the orientation of each
atom into its fundamental zone and projection into Rodrigues-Frank
space.  For materials with cubic symmetry, the fundamental zone (in
Rodrigues-Frank space) is a truncated octahedron~\cite{He:2008bc}, which permits a
straightforward mapping into RGB colour
space~\cite{{Albou:2010gv,VanBoxel:2005ih}}.  A drawback of the colour
scheme is that grains containing atoms whose fundamental orientations
lie close to the faces of the fundamental zone can have very different
colours.  This can be seen in the blue/beige grain (lower right) in
Figure~\ref{fig:render_orientation}.  Nevertheless, the colour scheme
conveys the relationship between the grain orientations well.

%\FloatBarrier
\section{Local Elastic Strain}
\label{sec:strain}
We have the optimal point-to-point correspondences between the actual atomic positions and the positions of the ideal structure.  As such, the local elastic strain is easily obtained, without any reference to an undeformed configuration at the beginning of the simulation.  First, we find the deformation gradient using a least-squares fit:
\begin{equation}
r = \min_{\set{A}} || \set{v}\set{A}^T - \set{w} ||_2
\label{eq:deformation_gradient}
\end{equation}
where $r$ is the residual term, \set{v} and \set{w} are $3\times N$ matrices containing the positions of the ideal positions and the optimally permuted actual positions respectively, and the deformation gradient, \set{A}, is the affine transformation which minimizes the residual term.  The residual term is equivalent to Falk and Langers $D^2_{\text{min}}$ term for identifying local irreversible shear transformations~\cite{Falk:1998hd}.  Prior to fitting the deformation gradient, the ideal and actual positions are translated such that the barycentre of each set lies at the origin, and scaled such that the mean vertex distance is 1, as in Equation (\ref{eq:rmsd_scaling}).  Although the deformation gradient obtained is scale-invariant, the scale factor can be used to recover the hydrostatic component.
The orientation and elastic strain matrices are obtained via a left-sided polar decomposition of the deformation gradient:
\begin{equation}
\set{P}\set{U} = \set{A}
\end{equation}
where \set{U} is an orthogonal right-handed matrix (the rotation matrix), and \set{P} is a symmetric matrix (the elastic strain matrix).  The choice of a left-sided polar decomposition is arbitrary, but we find the elastic strain in the same frame of reference preferable for comparison of strains across different grains.  In the case where \set{P} is not the identity matrix, \set{U} is not the same rotation found by minimizing the RMSD, since the addition of strain means we no longer have a rigid-body transformation.  The residual term in Equation~(\ref{eq:deformation_gradient}) could be used to determine the local structure instead of the RMSD, however, the elastic strains in MD simulations are typically less than $5\%$, and the extra degree of freedom provided by the strain matrix often results in highly-strained spurious structural identifications.

\begin{figure}[tp]
\centering
	\raisebox{-0.5\height}{\includegraphics[width=0.4\textwidth]{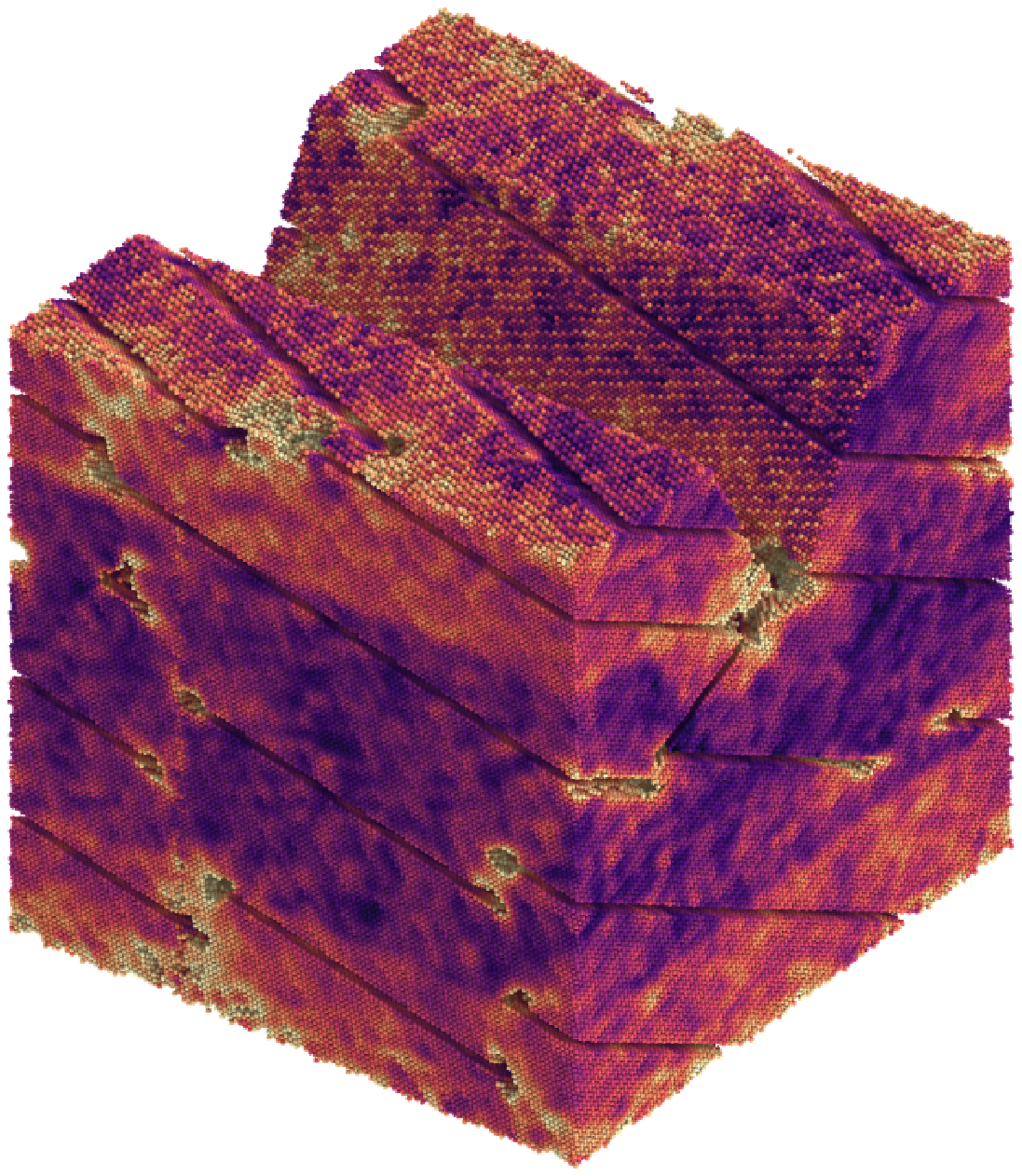}}
	\hspace{5mm}
    \raisebox{-0.5\height}{\includegraphics[height=0.25\textheight]{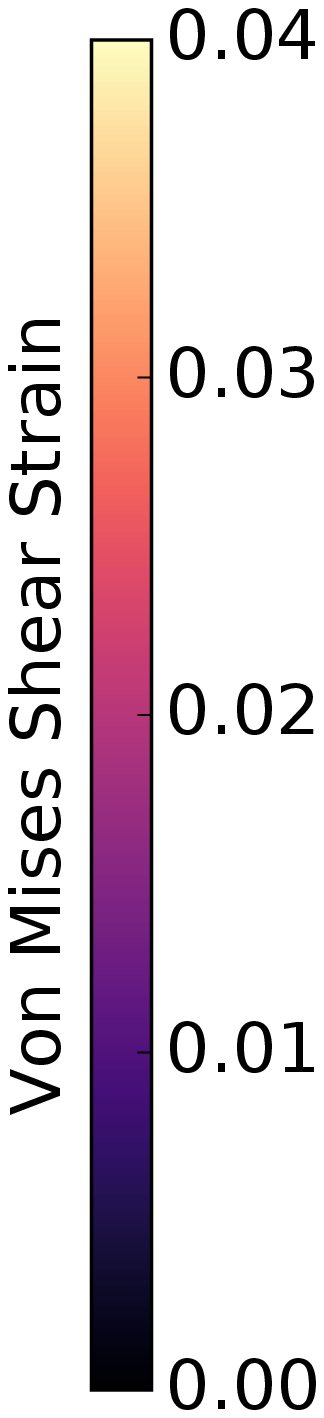}}
\caption{Local Von-Mises shear strain of FCC atoms in a single crystal $\text{CuPt}_3$ system, with $\text{RMSD}_{\text{max}} = 0.02$.  Regions of elastic strain occur around the wedge tip, close to surfaces and near dislocations.  Strain analysis requires quenching or time-averaging of positions, to avoid strains being dominated by atomic thermal displacements.}
\label{fig:render_strain}
\end{figure}

Figure~\ref{fig:render_strain} shows the Von-Mises shear strain for the FCC atoms in a $\text{CuPt}_3$ system.  The Von-Mises shear strain is given by:
\begin{equation}
\varepsilon_{\text{VM}} = \sqrt{\frac{3}{2}\sum_{ij}\set{P}_{ij}^2 - \frac{1}{2}\left(\sum_k \set{P}_{kk}\right)^2}
\end{equation}
% \begin{equation}
% \varepsilon_{\text{VM}} = \sqrt{\frac{1}{2}\left[(\set{P}_{11} - \set{P}_{22})^2 + (\set{P}_{22} - \set{P}_{33})^2 + (\set{P}_{33} - \set{P}_{11})^2 \right] + 3(\set{P}_{12}^2 + \set{P}_{23}^2 + \set{P}_{31}^2)}
% \end{equation}
where the sums are over the three coordinates.  This is the strain
equivalent of the Von Mises shear stress, defined in most solid
mechanics textbooks \cite{2011shigley}.
A wedge has been cut out along the length of the system, and a $5\%$ uniaxial strain has been applied transverse to the wedge tip direction.  Periodic boundary conditions have been applied along the wedge tip and strain directions.  The system was initially a single crystal (to avoid the strain field being dominated by grain boundaries) and has been annealed at 700K for 15000 steps of 5fs to allow stacking faults and dislocations to form.  Unfortunately, either quenching or time-averaging of atomic positions is necessary for strain analysis; since the strains are very small, they are dominated by thermal displacements at even moderate temperatures.  As such we have quenched the system using the FIRE~\cite{Bitzek:2006bw} method.  As expected, stress concentrations are seen near dislocation cores and at the tip of the wedge.  

%\FloatBarrier
\section{Implementation}
The core components of the PTM algorithm (RMSD optimization, convex
hull computation and graph canonization) can be computationally
expensive if not implemented carefully.  The RMSD optimization is solved using the library provided by D.L. Theobald~\cite{Theobald:2005cy, Liu:2010bc}, which is much faster than SVD based calculations.  Commonly used convex hull libraries, such as QHull~\cite{Barber:1996iv} and CGAL~\cite{cgal:hs-ch3-12b}, implement the QuickHull algorithm (with expected running time $O(n \log n)$) and are designed to achieve good performance on large point sets.  On small point sets such as ours theoretical running times are less important to performance than efficient memory allocation, layout and accesses.  As such, we have implemented a stack-allocated incremental convex hull algorithm~\cite{o1998computational} which, despite having a $O(n^2)$ theoretical running time, is
%mine:	0m22.450s
%qhull:	5m50.125s
%15.6 times faster
15 times faster than QHull for our point sets.  For graph canonization, we have implemented a single-embedding adaptation of Weinberg's algorithm~\cite{Weinberg:1966jy}, again with stack allocation.  Rather than representing the optimal Weinberg code as a list of edges, we have devised a hash function which uniquely represents all triconnected planar graphs as a 64-bit integer.  We have ensured that no hash collisions occur by generating all possible graphs of this type (up to 15 vertices) using the \texttt{plantri} program~\cite{Brinkmann:2007fw}.
The deformation gradients in Equation (\ref{eq:deformation_gradient}) are solved by precomputing a Moore-Penrose pseudoinverse matrix for each structure.  This reduces solution of a least-squares problem to matrix multiplication.  The polar decomposition of the deformation gradient is computed using a SVD library optimized for $3\times 3$ matrices~\cite{mcadams2011computing}.
For the benchmark system used in section \ref{sec:results_benchmark}, the main components account for the following percentages of the total running time: $46\%$ - convex hull construction, $20\%$ - canonical form calculation, $15\%$ - RMSD optimization (including scaling and translation of the points), and $9\%$ - strain calculation.  When using Euclidean ordering of neighbouring atoms, the method is capable of indexing over 100,000 atoms per second on a laptop computer using a single core, which is approximately $25\%$ slower than ACNA.  Topological neigbour ordering requires the computation of the Voronoi cell of each atom.  To do so, we calculate the Delaunay triangulation of a central atom and its 18 nearest neighbours, which allows for up to half the inner shell atoms in FCC and HCP lattices to be wrongly ordered by Euclidean distance.  We have implemented the Delaunay triangulation using the parabolic lifting map method of Edelsbrunner and Seidel~\cite{Edelsbrunner:1986fd} as it requires fewer intermediate simplices than the Bowyer-Watson~\cite{Bowyer:1981ed,Watson:1981dc} algorithm and consequently allows for a stack-allocated implementation.  Nonetheless, topological ordering requires significantly more computational effort than Euclidean ordering and increases the running time by a factor of 2.  The PTM source code is available online~\cite{PTMgithubrepository}.
%
% 32.31      4.51     4.51  2662110     1.69     2.93  get_convex_hull(int, double*, int, signed char (*) [3])
% 16.42     10.67     2.29 90116090     0.03     0.03  calculate_plane_normal(double const*, int, int, int, double*)
%  7.17     12.71     1.00 87453980     0.01     0.04  add_facet(double const*, int, int, int, facet_t*, double*)
% 27.75      8.38     3.87  1932264     2.00     2.00  canonical_form(int, signed char (*) [3], int, int, signed char*, signed char*)
%  7.46     11.71     1.04  6065230     0.17     0.30  normalize_vertices(int, double*, double (*) [3])
%  3.37     13.18     0.47  3485790     0.13     0.13  FastCalcRMSDAndRotation(double*, double*, double*, double, int, double)
%  2.01     13.46     0.28  3485790     0.08     0.08  InnerProduct(double*, int, double const (*) [3], double (*) [3], signed char*)
%
% 55.90 get_convex_hull(int, double*, int, signed char (*) [3])
% 27.75 canonical_form(int, signed char (*) [3], int, int, signed char*, signed char*)
% 12.84 optimizing rmsd (including scaling and translation)

\section{Conclusions}

We have presented a computationally efficient yet robust method for identifying the local structure in atomic-scale simulations.  The method is based on using the topology of the convex hull formed by the neighbours of an atom to construct a small set of candidate structures, the best of which is chosen based on the root-mean-square deviation of the positions from their expected positions.  In case of local crystalline symmetry, the method also identifies the local orientation of the crystalline axes, and optionally the elastic strain tensor.  At low to moderate temperatures, the performance of the method is very similar to preexisting methods such as Adaptive Common Neighbour Analysis, but at higher temperatures it is more robust, and can with high reliability identify the structure of a crystalline phase for all temperatures up to the melting point.

\section*{Acknowledgments}

We would like to thank E. A. Lazar for fruitful discussions of the manuscript. 
We gratefully acknowledge funding from The Danish National
Research Foundation's Center for Individual Nanoparticle Functionality
(DNRF54).

%\FloatBarrier
\appendix
\setcounter{table}{0}
\renewcommand{\thetable}{A\arabic{table}}

\section{Symmetrically Inequivalent Triangulations}
\label{sec:app_unique}
In section~\ref{sec:convexhulls} we described how a convex hull can have multiple triangulations of its surface.  Here we describe how the symmetrically unique triangulations can be generated, which results in a smaller number of templates to match against a structure.

\begin{figure}[tp]
\centering
\begin{minipage}{.19\textwidth}
  \centering
  \includegraphics[width=1.0\textwidth]{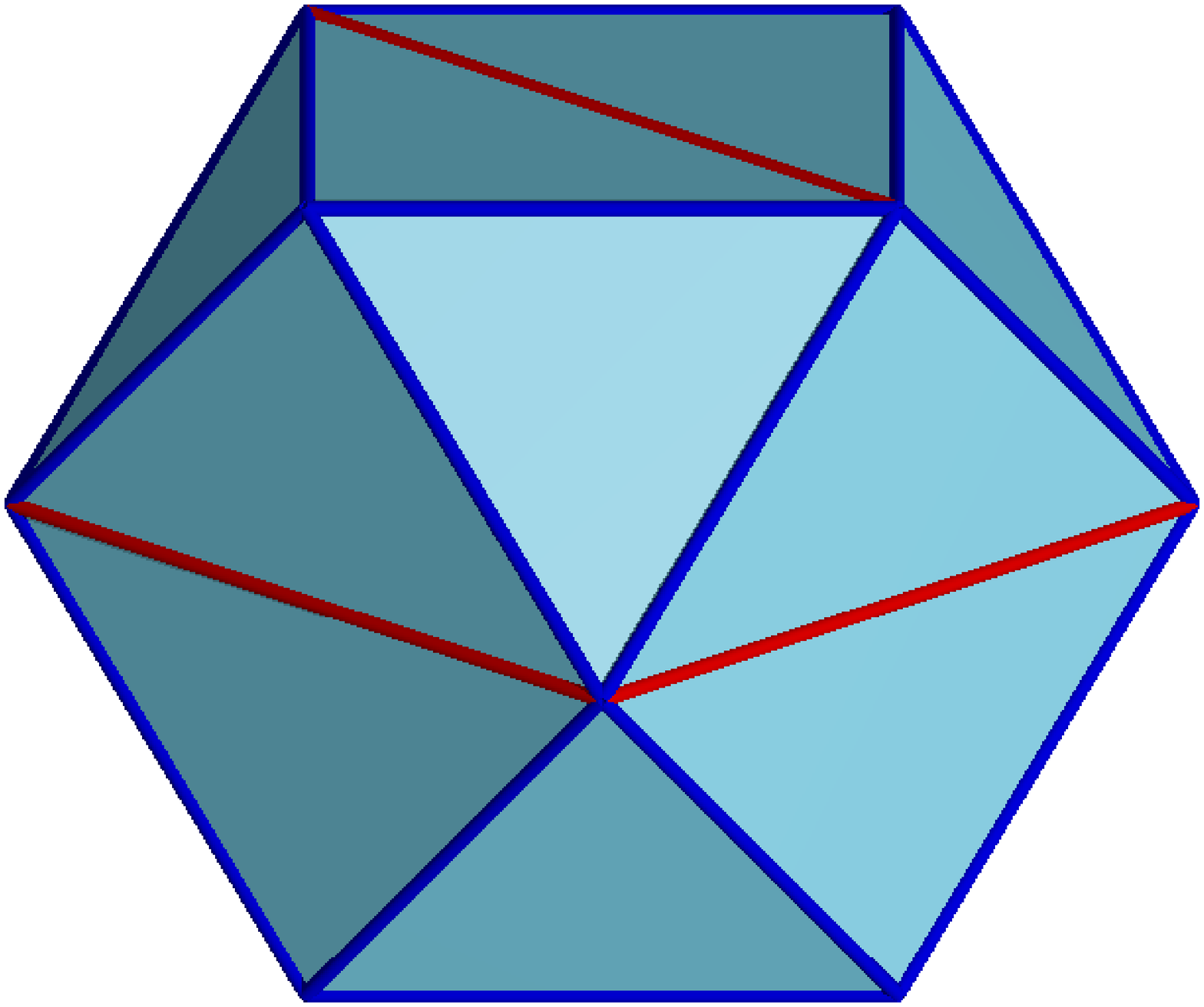}\\
  FCC
\end{minipage}
\hspace{10mm}
\begin{minipage}{.19\textwidth}
  \centering
  \includegraphics[width=1.0\textwidth]{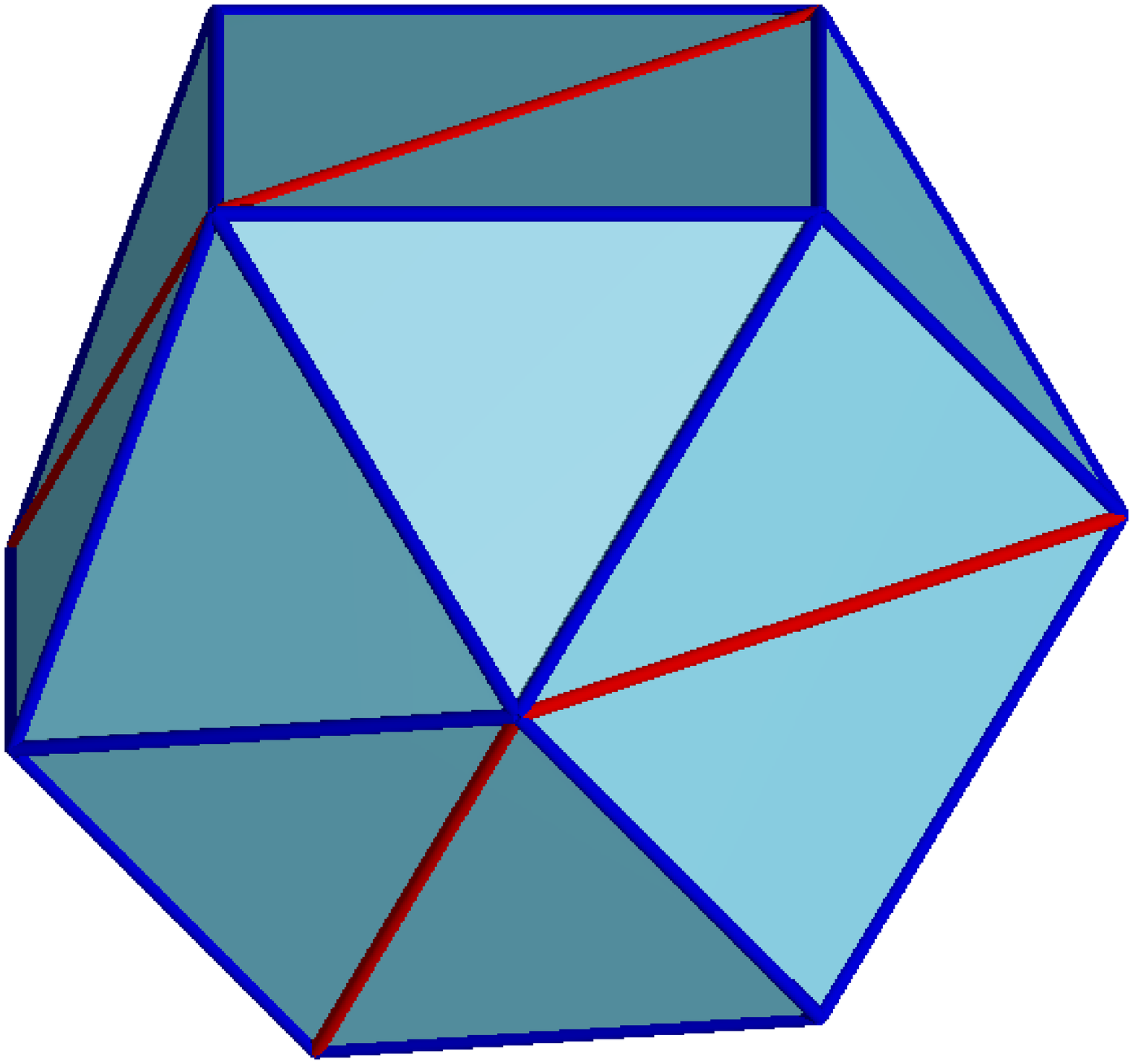}\\
  HCP
\end{minipage}
\caption{The two different edge types in the convex hull graphs of the FCC and HCP structures; edges which cross a square facet (red) and edges which do not (blue).}
\label{fig:edge_colours}
\end{figure}

The SC and ICO structures are simple cases.  They both have single convex hull triangulations.  Whilst the graph of their convex hulls have 24 and 60 orientation-preserving automorphisms respectively, these automorphisms are symmetrically equivalent.
The FCC and HCP cases are slightly more complex.  We can generate the unique triangulations by assigning colours to the edges in the convex hull graph.  There are two edge lengths in the convex hull; edges which cross a square facet and edges which do not.  This is shown in Figure~\ref{fig:edge_colours}.  The unique graphs are now the unique edge-colour-preserving and orientation-preserving isomorphisms.  The symmetrically unique automorphisms can be determined either by finding the unique RMSD values, or by colouring the facets by type (equilateral triangle or half a square facet) and finding the unique orientation-preserving isomorphisms of the dual graph.

The unique BCC triangulations can be found by colouring the vertices by shell number, and finding the unique vertex-colour-preserving and orientation-preserving isomorphisms.  The unique automorphisms can be found either by finding the unique RMSD values, or by colouring the facets according to the number of vertices in each shell and finding the unique orientation-preserving isomorphisms of the dual graph.
Whilst we assign colours to edges and vertices during template generation, no colours are assigned during template matching.  Colour assignment requires prior knowledge of facet types and, in the BCC case, the neighbour vertex shell numbers.  However, the purpose of template matching is to accurately determine the structure.  The lack of assumptions about facet types and shell numbers results in robust structural identifications even in highly distorted structures.

%\FloatBarrier
\section*{References}
\bibliographystyle{iopart-num}
\renewcommand\refname{}
\bibliography{newrefs}

\end{document}